\documentclass[
reprint,
superscriptaddress,
amsmath,amssymb,
aps,
pra,
]{revtex4-2}

\usepackage{graphicx}
\usepackage{dcolumn}
\usepackage{bm}
\usepackage{hyperref}
\usepackage{graphicx} 
\usepackage{xcolor}
\usepackage{framed}

\DeclareMathOperator{\Tr}{Tr}
\def  \hrho  {\hat{\rho}}
\def  \hmu   {\hat{\mu}}

\def  \calc  {\mathcal{C}}
\def  \loss  {\mathcal{L}}
\def  \bS    {{\bm S}}
\def  \bC    {{\bm C}}
\def  \Strain  {{\bm S}_\mathrm{train}}
\def  \Stest   {{\bm S}_\mathrm{test}}
\def  \acc   {\mathrm{ACC}}
\def  \prec  {\mathrm{PREC}}
\def  \rec   {\mathrm{REC}}
\def  \mse   {\mathrm{MSE}}
\def  \rmse  {\mathrm{RMSE}}
\def  \Er    {\mathrm{E}}

\def  \bMA   {{\bm M}_\mathrm{A}}
\def  \bMB   {{\bm M}_\mathrm{B}}
\def  \bMC   {{\bm M}_\mathrm{C}}
\def  \bMD   {{\bm M}_\mathrm{D}}

\begin{document}


\title{Learning entanglement from tomography data: contradictory measurement importance for neural networks and random forests}

\author{Pavel Bal\'{a}\v{z}}
\affiliation{FZU - Institute of Physics of the Czech Academy of Sciences, Na Slovance 1999/2, 182 00 Prague, Czech Republic}

\author{Mateusz Krawczyk}
\affiliation{Institute of Theoretical Physics, 
Wroc{\l}aw University of Science and Technology,
Wybrze\.{z}e Wyspia\'{n}skiego 27,
50-370 Wroc{\l}aw, Poland}

\author{Jaros\l{}aw Paw\l{}owski}
\affiliation{Institute of Theoretical Physics, 
Wroc{\l}aw University of Science and Technology,
Wybrze\.{z}e Wyspia\'{n}skiego 27,
50-370 Wroc{\l}aw, Poland}

\author{Katarzyna Roszak}
\affiliation{FZU - Institute of Physics of the Czech Academy of Sciences, Na Slovance 1999/2, 182 00 Prague, Czech Republic}

\begin{abstract}
We study the effectiveness of two distinct machine learning techniques, neural networks and random forests, in the quantification of entanglement from two-qubit tomography data.
Although we predictably find that neural networks yield better accuracy, we also find that 
the way that the two methods reach their prediction is starkly different. This is seen by the measurements which are the most important for the classification. Neural networks follow the intuitive prediction that measurements containing information about non-local coherences are most important for entanglement, but random forests signify the dominance of information contained in occupation measurements. This is because occupation measurements are necessary for the extraction of data about all other density matrix elements from the remaining measurements.
The same discrepancy does not occur when the models are used to learn entanglement
directly from the elements of the density matrix, so it is the result of the 
scattering of information and interdependence of measurement data.
As a result, the models behave differently when noise is introduced to various
measurements, which
can be harnessed to obtain more reliable information about 
entanglement from noisy tomography data.
\end{abstract}

\maketitle

\section{Introduction}
Pure state entanglement is in principle straightforward to determine, since there is a direct correspondence between the amount of entanglement between subsystems and the purity of the density matrix
of either subsystem. This is obvious, since a measure of entanglement, analogous to entanglement entropy,
but based on linear entropy instead of von Neumann entropy, is a good and reliable entanglement measure.
Experimental determination of purity cannot be performed directly and quantum state tomography is required, 
but it is enough to measure the state of the smaller of the entangled subsystems to describe pure state entanglement.

For mixed states, even theoretical determination of entanglement becomes challenging and it's experimental studies
require quantum state tomography (QST) of the whole density matrix~\cite{James2001} followed by the calculation of one of the more numerically accessible 
entanglement measures. Since quantum state tomography requires $\mathcal{N}^2-1$ measurements, where $\mathcal{N}$ is the dimension of the whole system, the number of required measurements grows rapidly with system size (i.e. $\mathcal{N}=2^\mathrm{\#\,qubits}$) and becomes unmanageable for pretty small systems. 
For larger systems quantum state reconstruction can be effectively (with learned positivity constraint) performed using deep neural networks (NNs)~\cite{Torlai2018,Xin2019,Melkani2020,Shahnawaz2021,Quek2021,Kotuny2022,Schmale2022,Ma2024}, that can converge much faster~\cite{Kotuny2022} than the standard QST, even for noisy measurement data~\cite{Ma2024}.

The number of measurements required for full state tomography of the smallest possible system that can
be entangled, namely a two qubit system, is already non-negligible. Although it is standard in optics,
in many systems it is not possible to perform the required $4^2-1=15$ two-qubit measurements, especially that 
a very good control of the measurement basis is required to obtain all of the information about the
two-qubit density matrix. Thus a situation where not all of the measurements can be performed and 
estimation of entanglement must be performed from incomplete or noisy data is very natural.

Machine learning (ML) has proven its usability in recognizing entanglement with (classical) deep NNs~\cite{Krawczyk2024,Pawlowski2024,Taghadomi2024,Chen2022,Asif2023,urena2024}, or non-neural models~\cite{Lu2018}, e.g., using ensemble learning~\cite{Hiesmayr2021,Goes2024,wang2024}. Ensemble learning is a powerful paradigm in ML in which multiple models are combined to enhance overall predictive performance. This technique capitalizes on the diversity of base learners, leading to a more generalized and reliable model. In scientific applications, ensemble learning has demonstrated significant benefits, from reducing overfitting to enhancing the stability of predictions, making it an indispensable tool in the modern ML toolkit~\cite{Ganaie2022}.
Random Forest (RF)~\cite{Breiman2001} is one of the best known examples of ensemble methods that construct multiple decision trees (DTs)~\cite{Furnkranz2010,breiman2017classification} using the principle of bagging~\cite{Breiman1996} and selecting random features. 

In the last decade, the revolution in AI has been driven by the rapid development of deep NNs~\cite{LeCun2015,unet}, which have demonstrated remarkable success in tasks like image processing and natural language analysis~\cite{vaswani2017}. 
Unlike DTs or dimensionality reduction techniques, NNs take a different approach to machine learning, with a structure inspired by biology but also physical models~\cite{Gibney2024}. While they allow for the training of highly complex models far beyond traditional ML methods~\cite{belkin2021}, their interpretability becomes a challenge. Nevertheless, efforts are underway to develop techniques that make NNs more explainable~\cite{Longo2024}, especially in science where understanding how a model works is as important as the quality of its predictions.

In this work, we study the possibility of obtaining reliable information about two-qubit entanglement from quantum tomography measurements
on the basis of incomplete or perturbed experimental data. To this end, we employ two distinct machine learning (ML) techniques, the NN model and the RF model.
We predictably find that the NN model is better at evaluating entanglement from the full
set of measurement data.
We further evaluate
the importance of the different measurments both directly
and by introducing noise for a chosen measurement.
Here we consistently find that that the NN model behaves intuitively, meaning that 
measurements containing information about non-local coherences are the most relevant,
followed by those with information about local coherences, and occupations are least 
relevant. This behavior is in agreement with feature importance displayed when
concurrence is calculated from tomography data, without ML (as quantified by the Shapley values). 

The RF model displays completely opposite behavior. It signifies that measurements that
quantify occupations are the most important, while those that depend on the non-local coherences
are least important. This discrepancy occurs because the data about the system state is 
provided as a set of measurements, in which the information is dispersed. A similar calculation performed using density matrix elements
directly showed no discrepancy and the feature importance followed the intuitive order of
importance for entanglement: non-local coherences, local coherences, occupations. 

This means that under circumstances where some data may be unreliable, predictions 
made by substantially different ML techniques are likely to vary and the better choice 
of method depends on the type of problem under study. NN models act in a more direct way
and for the specific problem of entanglement quantification perform better for a full set 
of tomography data. The inner working of RF models is more intricate and they are much more 
dependent on the interdependence of the measurement data, thus shifting importance onto occupations.
This is an advantage, since occupations are typically easier to measure than coherences.

The paper is organized as follows. In Secs \ref{sec2} and \ref{sec3} we provide the definition and discussion of the concurrence and introduce basic notions about two-qubit
quantum state tomography, respectively. In Sec.~\ref{sec4} we discuss the datasets that are used in the study. Sec.~\ref{sec5} introduces the two types of machine learning models. Sec.~\ref{sec6} contains a description of the results and their discussion. Sec.~\ref{sec8} concludes the paper.


\section{Concurrence\label{sec2}}

The classification of two-qubit entanglement is reasonably straightforward even for mixed states, since there
exists a formula for the (numerical) calculation of Entanglement of Formation (EoF) \cite{Wootters1998} directly form the density 
matrix (no minimization is required and bound states \cite{horodecki97,horodecki98} do not exist). In the following we use the concurrence as the measure of entanglement. The concurrence is unambigously related to EoF \cite{Wootters1998},
while it is a bit easier to use. 

The concurrence is a unique measure that exists only for two-qubit systems
(contrarily to EoF, which can be defined for a system of any size). It is defined as
\begin{equation}
	C(\rho)=\mathrm{max}(0, \lambda_1 - \lambda_2 - \lambda_3 - \lambda_4),
	\label{eq:concurrence}
\end{equation}
where $\lambda_1 \geq \lambda_2 \geq \lambda_3 \geq \lambda_4$ are the eigenvalues of the matrix 
\begin{equation}
	R = \sqrt{\sqrt{\rho}(\sigma_y \otimes \sigma_y)\rho^*(\sigma_y \otimes \sigma_y)\sqrt{\rho}},
\end{equation}
in decreasing order.  Here, $\sigma_y$ denotes the appropriate Pauli matrix and $\rho^*$
is the complex conjugate of the two-qubit density matrix. 
Concurrence is an entanglement monotone, which ranges between $0$ and $1$.  A concurrence value $C(\rho) = 0$ 
signifies that the state $\rho$ is separable (there is no entanglement between the qubits), while 
$C(\rho) = 1$ indicates a maximally entangled state.

Let us now look at a generic two-qubit density matrix
written in the standard separable basis $\{|00\rangle,|01\rangle,|10\rangle,|11\rangle\}$,
\begin{equation}
\label{dm}
\left(
\begin{array}{cccc}
 \color{red}\rho_{00,00} & \rho_{00,01} & \rho_{00,10} & \color{blue}\rho _{00,11} \\
 \rho^\ast_{00,01} & \color{red}\rho_{01,01} & \color{blue}\rho _{01,10} & \rho _{01,11} \\
 \rho^\ast_{00,10} & \color{blue}\rho^\ast_{01,10} &\color{red}\color{red} \rho_{10,10} & \rho_{10,11} \\
 \color{blue}\rho^\ast_{00,11} & \rho^\ast_{01,11} &\rho^\ast_{10,11}& \color{red}\rho_{11,11}\\
\end{array}
\right)
\end{equation}
with
$\rho_{11,11} = 1-\rho _{00,00}-\rho_{01,01}-\rho_{10,10}$.
We color-coded the elements of the density matrix, marking the diagonal elements (occupations) red, the non-local coherences blue,
and leaving the other off-diagonal elements black, for clarity in the subsequent discussion. 
The importance of the different elements in this basis for entanglement is not uniform. For instance, a state that has non-zero elements only on the diagonal
is a statistical mixture and can contain only classical correlations. Non-zero non-local coherences are, on the other hand, a necessary condition 
for two-qubit entanglement, since they contain information about the phase relations between the qubits. This is immediately evident when studying
X-states \cite{yu07,MENDONCA201479}, but is true for any two-qubit state. The interplay between the 
non-local coherences with each other and with 
other off-diagonal elements plays a significant role for both the existence of entanglement as well as it's quantity. The magnitude of the different off-diagonal
elements with respect to diagonal elements contains information about the purity of the state and as such is also critical for entanglement.

\section{Two-qubit quantum state tomography \label{sec3}}

In the following, we use the projective-measurement tomography scenario as in the original proposal of Ref.~\cite{James2001}.
To determine the state of a single qubit via quantum state tomography, only three distinct measurements are 
required, which are typically chosen to correspond to the three Pauli operators. For two qubits, the minimum
required number of measurements is $15$,  but in the following we always include the superfluous $16$-th measurement for the sake of symmetry.
In general, the outcome of a projective measurement can be written as~\cite{James2001}
$m = {\cal N} \Tr\left\{ \hrho\, \hmu \right\}$,
where $\hat{\rho}$ is the density matrix, $\hat{\mu}$ is the projection operator, and ${\cal N}$ is a constant of proportionality which can be determined from the data.
For two qubits, the outcomes of joint measurements required for tomography can be expressed as
\begin{equation}
\label{Eq:measurements}
    m_{ij} = {\cal N} \Tr\left\{ 
    \hrho \left( \hmu_i \otimes \hmu_j \right) 
    \right\},
\end{equation}
where $\hmu_i$ and $\hmu_j$ ($i,j = 0, 1, 2, 3$) are projectors on a given single qubit state. 
Note that this formulation limits the measurements to measurements in separable bases.

The choice of measurements required to obtain full information about a quantum state is not unique.
In the following, we use the set of measurements proposed in Ref.~\cite{James2001},
where the single qubit measurements, $\hmu_i =|i\rangle\langle i |$, correspond to projections on states
\begin{equation}
    \label{eq:basis}
    |0\rangle,|1\rangle,
    |2\rangle = \frac{1}{\sqrt{2}}\left(|0\rangle+|1\rangle\right),
    |3\rangle =\frac{1}{\sqrt{2}}\left(|0\rangle-i|1\rangle\right).
\end{equation}

Below we provide expectation values for each measurement, expressed with the help of the elements of the density 
matrix (\ref{dm}) and retaining the color coding in order to facilitate the discussion on how the information about the density matrix
is distributed within the measurements.
Throughout the paper we group the measurements into four blocks as follows
\begin{equation}
    \begin{split}
        \bMA &= \{m_{00}, m_{01}, m_{10}, m_{11}\}\, \\   
        \bMB &= \{m_{02}, m_{03}, m_{12}, m_{13}\}\,,\\   
        \bMC &= \{m_{20}, m_{21}, m_{30}, m_{31}\}\,, \\  
        \bMD &= \{m_{22}, m_{23}, m_{32}, m_{33}\}\,.     
    \end{split}
    \label{eq:blocks}
\end{equation}
The blocks contain information about different types of density matrix elements as is evident below.

Block $\bMA$ contains all measurement outcomes that can be obtained in the $|ij\rangle$, $i,j=0,1$ two-qubit basis
and thus the measurements only yield information about the diagonal elements of the matrix (\ref{dm}),
\begin{subequations}
\label{bloka}
\begin{eqnarray}
m_{0,0}&=&\color{red}\rho _{00,00},\\
m_{0,1}&=&\color{red}\rho _{01,01},\\
m_{1,0}&=&\color{red}\rho_{10,10},\\
m_{1,1}&=&\color{red}\rho_{11,11}.
\end{eqnarray}
\end{subequations}

Blocks $\bMB$ and $\bMC$ contain measurement outcomes, where one of the qubits is measured in the $\{|0\rangle,|1\rangle\}$, basis while the other is measured
with respect to one of the other two states used in the tomography [see Eqs.~(\ref{eq:basis})].
Thus the dependence of outcomes on the elements of the density matrix from block $\bMB$ can be obtained by exchanging qubit indices in the outcomes 
from block $\bMC$ and vice versa. Note that this symmetry does not translate into any of the measurements being superfluous, since a density matrix does 
not have to contain any symmetry with respect to the exchange of qubits.
The explicit formulas for blocks $\bMB$ and $\bMC$ are given by 
\begin{subequations}
\label{blokb}
\begin{eqnarray}
m_{0,2}&=&\frac{1}{2} \left({\color{red}\rho _{00,00}}+{\color{red}\rho _{01,01}}+2 \Re\left(\rho _{00,01}\right)\right),\\
m_{0,3}&=&\frac{1}{2} \left({\color{red}\rho_{00,00}}+{\color{red}\rho_{01,01}}+2\Im\left(\rho _{00,01}\right)\right),\\
m_{1,2}&=&\frac{1}{2} \left({\color{red}\rho_{10,10}}+{\color{red}\rho_{11,11}}+2\Re\left(\rho _{10,11}\right)\right),\\
m_{1,3}&=&\frac{1}{2} \left({\color{red}\rho_{10,10}}+{\color{red}\rho_{11,11}}+2\Im\left(\rho _{10,11}\right)\right),
\end{eqnarray}
\end{subequations}
and
\begin{subequations}
\label{blokc}
\begin{eqnarray}
m_{2,0}&=&\frac{1}{2} \left({\color{red}\rho_{00,00}}+{\color{red}\rho _{10,10}}+2\Re\left(\rho _{00,10}\right)\right),\\
m_{2,1}&=&\frac{1}{2} \left({\color{red}\rho_{01,01}}+{\color{red}\rho _{11,11}}+2\Re\left(\rho _{01,11}\right)\right),\\
m_{3,0}&=&\frac{1}{2} \left({\color{red}\rho _{00,00}}+{\color{red}\rho_{10,10}}+2\Im\left(\rho _{00,10}\right)\right),\\
m_{3,1}&=&\frac{1}{2} \left({\color{red}\rho _{01,01}}+{\color{red}\rho _{11,11}}+2\Im\left(\rho _{01,11}\right)\right),
\end{eqnarray}
\end{subequations}
respectively.
These measurements contain information about the local coherences, but not the non-local coherences.
This means that the information contained in the two blocks is insufficient to distinguish between an entangled and a separable state. 

The last block, $\bMD$, is the block that contains information about the critical non-local coherences,
\begin{subequations}
\label{blokd}
\begin{eqnarray}
m_{2,2}&=&\frac{1}{4} \left(2 \Re\left(\rho _{00,01}+\rho_{00,10}+{\color{blue}\rho_{00,11}}+{\color{blue}\rho_{01,10}}\right.\right.\nonumber\\
&&+\left.\left.\rho_{01,11}+\rho_{10,11}\right)+1\right),\\
m_{2,3}&=&\frac{1}{4}\left(2 \Im\left(\rho _{00,01}+{\color{blue}\rho _{00,11}}-{\color{blue}\rho_{01,10}}+\rho _{10,11}\right)\right.\nonumber\\
&&+\left.2\Re\left(\rho _{00,10}+\rho _{01,11}\right)+1\right),\\
m_{3,2}&=&\frac{1}{4} \left(2\Im\left(\rho _{00,10}+{\color{blue}\rho _{00,11}}+{\color{blue}\rho _{01,10}}+\rho _{01,11}\right)\right.\nonumber\\
&&+\left.2\Re\left(\rho _{00,01}+\rho _{10,11}\right)+1\right),\\
m_{3,3}&=&\frac{1}{4} \left(2\Im\left(\rho _{00,01}+\rho_{00,10}+\rho_{01,11}+\rho _{10,11}\right)\right.\nonumber\\
&&+\left.2\Re\left({\color{blue}\rho _{01,10}}-{\color{blue}\rho _{00,11}}\right)+1\right).
\end{eqnarray}
\end{subequations}
Note that the measurement outcomes also depend on other off-diagonal elements, thus it is impossible to determine the magnitude of the non-local coherences without the other measurements.  


\section{Dataset \label{sec4}}

We generate a set of two-qubit density matrices by using different random-sampling methods. 
The main idea is to generate random states using the quantum circuits approach \cite{Pawlowski2024}, but in order to have a well-diversified ensemble, we also use a technique based on sampling from the uniform Haar measure \cite{mezzadri2006}. Additionally, to increase the number of maximally entangled states, we specifically include $20 000$ of them. In total, we generate $460 000$ density matrices for the training dataset
and $46 000$ matrices for the test set. All density matrices are labeled with the concurrence as defined by Eq.~({\ref{eq:concurrence}}).
We then transform the density matrices into feature vectors, that contain the outcomes of the two-qubit
measurements, Eq.~(\ref{Eq:measurements}). Feature vectors are composed of measurements together with respective labels $C$ representing the concurrence values form the training dataset $\Strain$.
During the generation process, we specifically restrict our sampling methods to generate a dataset balanced in the number of separable and entangled states. Due to this, the number of samples generated by various methods is adjusted to obtain a similar number of states with $C(\rho) < \tau$ and $C(\rho) >= \tau$, with $\tau=10^{-6}$, which is the chosen threshold value between separable and entangled states. More details regarding generation methods with a similar approach can be found in Ref.~\cite{Pawlowski2024}.
In the end, our training dataset $\Strain$ consists of about $53\%$ entangled states and $47\%$ separable states.
\begin{figure}[tb]
	\centering
	\includegraphics[width=0.9\columnwidth]{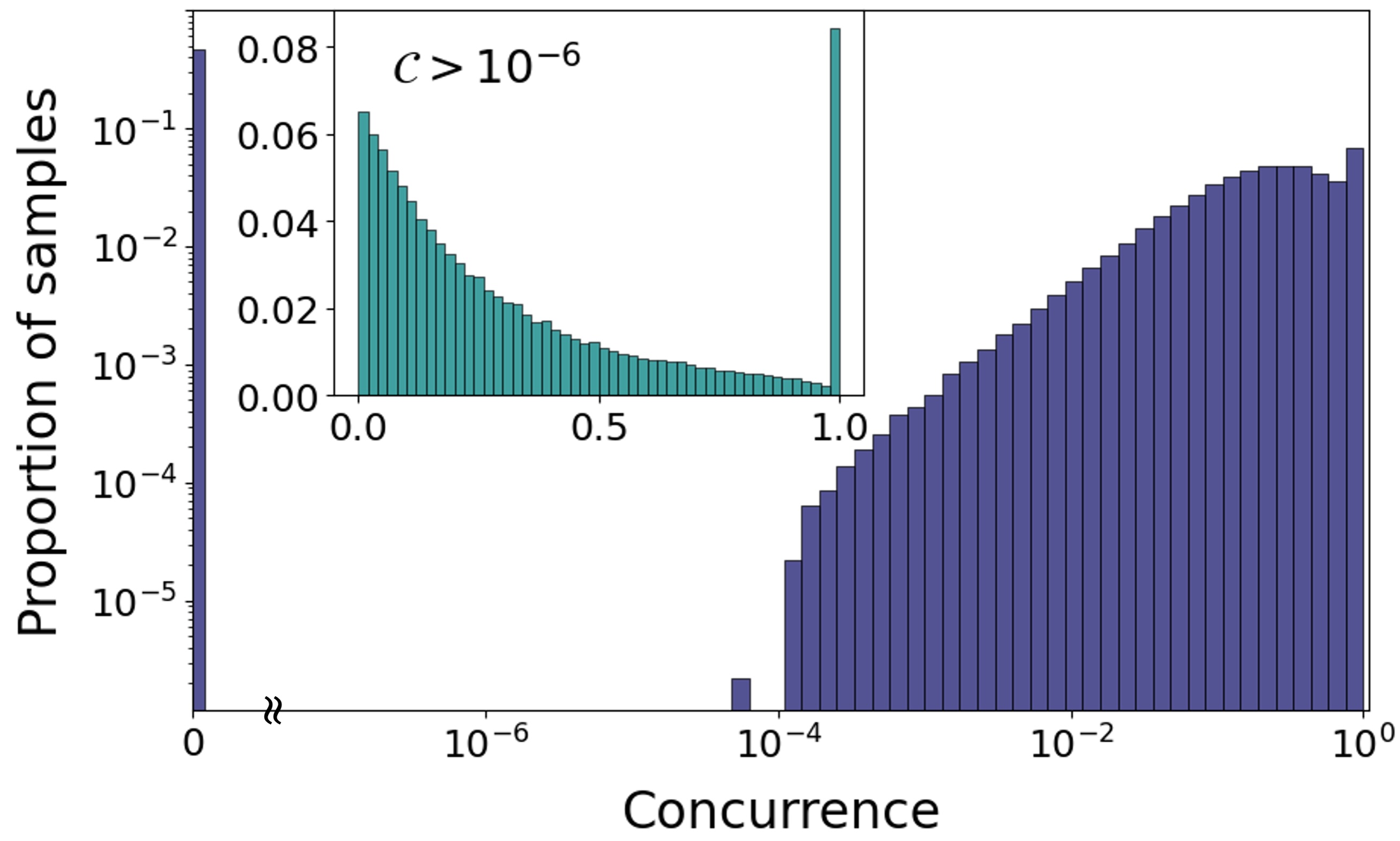}
	\caption{Proportion of samples in the training dataset with the concurrence within a given range. The main figure shows data for all samples. In the inset we show just entangled samples, i.e. with the concurrence $C > 10^{-6}$.}
	\label{fig:conc}
\end{figure}

We show the distribution (histogram) of the concurrence values $C$ across the training set $\Strain$ in Fig.~\ref{fig:conc}. The bars show the proportion of samples with $C$ within a given range. The main figure shows the whole range of the $C$ values in a logarithmic scale. We observe two regions corresponding to separable and entangled states. In the inset of Fig.~\ref{fig:conc} we see the distribution of concurrence of the entangled states only (with $\calc > 10^{-6}$). 
The number of samples decreases with increasing concurrence, however, there is a peak for samples with $C(\rho) \simeq 1$, i.e. maximally entangled, reflecting the presence of artificially included randomized Bell states. 
The test set $\Stest$ follows the same distribution.


\section{Models \label{sec5}}
\label{sec:models}

We examine two predominant types of machine learning models utilized in data classification and regression: an ensemble method (RF), and a deep NN (multilayer perceptron, MLP), and elucidate their respective advantages and disadvantages in the context of quantum state recognition.

\subsection{Ensemble methods}


Ensemble models are primarily divided into two categories: bagging~\cite{Breiman1996} and boosting~\cite{schapire1990strength,breiman1998arcing}.
One of the most popular bagging ensembles are Random Forests (RFs)~\cite{Breiman2001}. 
RFs utilize an ensemble of DTs (as base learners), which are a supervised learning technique that generates predictions by recursively partitioning data into subsets according to feature values~\cite{breiman2017classification}. In this study, we utilize the RF models for the classification of measurement vectors into two categories: {\em separable} (negative) or {\em entangled} (positive). Further details on the training of RFs and DTs, as well as their hyperparameters, can be found in Appendix~\ref{apx:rf}. 

\subsection{Deep neural networks}

\begin{figure}[bt]
\includegraphics[width=8.8cm]{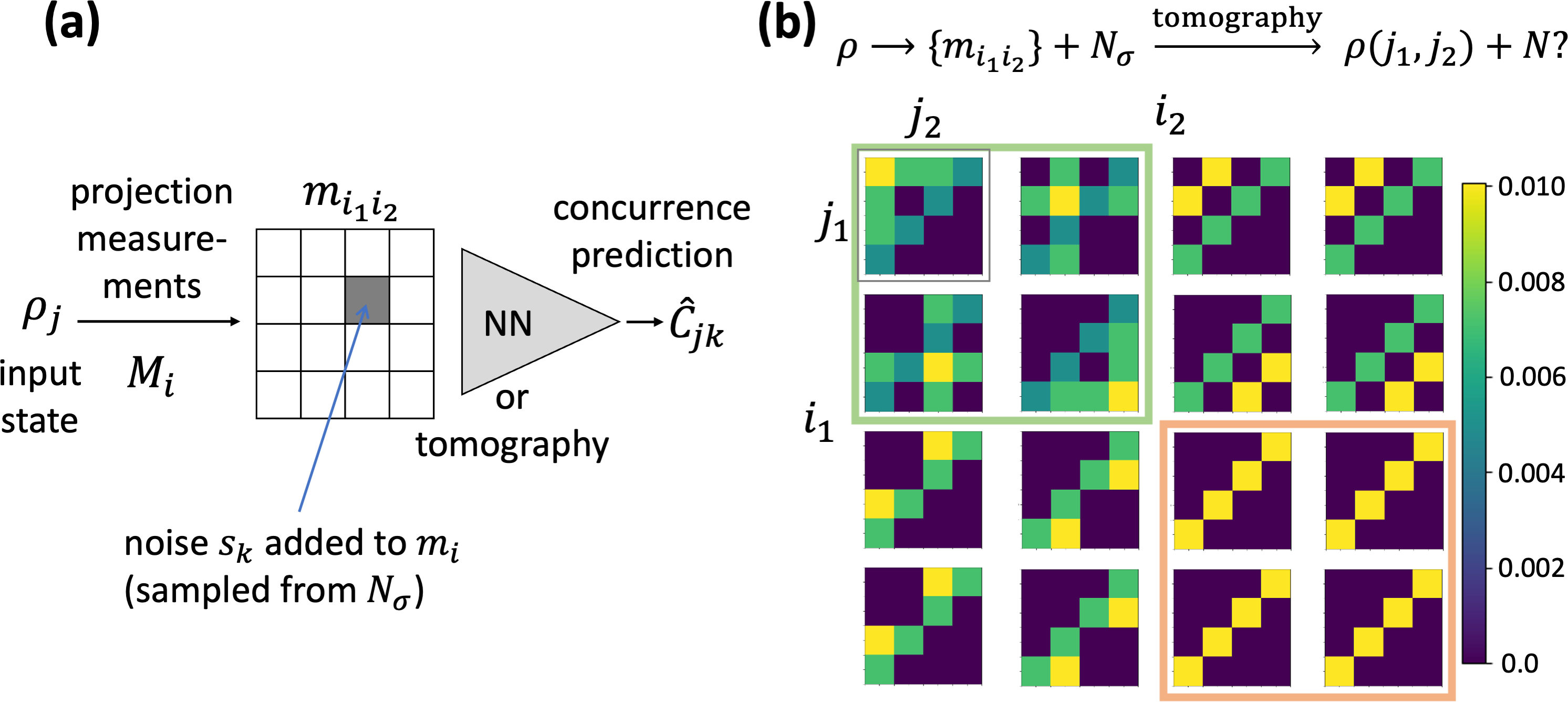}
\caption{Measurement perturbations: (a) methodology, (b) perturbation propagation within the quantum tomography: two distinct blocks of measurements are marked by green ($\bMA$, upper left) and orange ($\bMD$, lower right) boxes.}
\label{fig:importance_sheme}
\end{figure}

The second type of ML model employed is the NN MLP model.
MLPs are supervised models that have been successfully applied to both classification and regression tasks. 
In this work, we employed an MLP to accurately predict the value of concurrence $C$ 
using measurement values as inputs, as presented in Fig.~\ref{fig:importance_sheme}(a).
Our MLP architecture comprises just two \textit{fully-connected} hidden layers, each containing $128$ units (neurons), and utilizing a ReLU activation function~\cite{Nair2010}. 
Finally, it has a single output unit with a linear activation function that predicts the $C$ value.
We used a linear output activation, although a sigmoid function may seem more natural. However, combining mean square error (MSE) as the loss function with a sigmoid activation often leads to optimization issues, as the resulting cost surface becomes non-convex. To ensure that the predictions for $C$ remained within its natural range, we applied additional clipping.
We limited the model to only 2 hidden layers, which proved sufficient for making 2-qubit predictions. However, we also trained a much larger model consisting of 10 hidden layers, but it achieved comparable results. 

The MLP is trained using the training dataset, $\Strain$, (of size $N$) 
using a standard MSE loss function,
\begin{equation}
    \loss(\Strain) \equiv 
    \mse(\bC, \hat{\bC}) = 
    \Er_j\!\left[(C_j - \hat{C}_j)^2\right],
\end{equation}
where $C_j$ is the ground truth value for concurrence of a given state $j$, while $\hat{C}_j$ is the 
the concurrence estimated by the NN model. 
Moreover, $\Er_j[X_j]=(1/N)\sum_{j=1}^{N}X_j$ denotes a standard average over an index $j$.

\subsection{Measurements perturbation in QST}
To improve the understanding of the measurements' impact on concurrence prediction, we also analyze subsequent measurements' impact on QST by introducing perturbations to the standard tomography.
Fig.~\ref{fig:importance_sheme} (b) shows, how measurement perturbations propagate through quantum tomography on a reconstructed state $\rho$. There we present average errors in the reconstructed state $\rho(j_1,j_2)$, with density matrix elements numbered by $j_1$ and $j_2$, upon the addition of a Gaussian noise $N_\sigma$ with $\sigma=0.01$ to subsequent measurements $m_{i_1,i_2}$. 
In this way we obtain information on how given measurements impact $\rho$ element reconstruction. We observe that various measurements play different roles during the reconstruction.
For example the $\bMD$ block (orange box) explains only the coherences of the density matrix $\rho$. On the other hand, block $\bMA$ (green box) has impact on all the elements of $\rho$,
which is the consequence of the fact that the diagonal elements which are measured by this
block have to be used to infer the off-diagonal elements from measurement blocks $\bMB$
and $\bMC$. These in turn are critical in determining the inter-qubit coherences. 
Block $\bMD$ measurements do not impact any other elements of the density matrix than the 
non-local coherences.

\section{Results \label{sec6}}

In this section, we conduct an evaluation of our models utilizing the test dataset, $\Stest$, comprising $46000$ samples that were not incorporated during the training phase. Subsequently, we apply interpretability methods to elucidate the most salient aspects of the models' predictions.

\subsection{Predictive power of the ML models}

We start with ensemble models (RFs listed in Tab.~\ref{tab:rf_models} in Appendix~\ref{apx:rf}).
For these models, we compute the accuracy, which is defined as
\begin{equation}\label{eq:rf_acc}
    \acc_\mathrm{RF}\left(\Stest \right) = 
    \Er_j \left[ 1 - \left| P_j - \hat{P}_j \right| \right]\,,
\end{equation}
where $P_j$ is the ground truth value for the predicted sample class (binary: separable or entangled), while
$\hat{P}_j$ is the model's prediction.
In the scenario where $\hat{P}_j = 0$ holds, the model infers that the observations pertain to two separable qubits, thereby classifying them as the negative class. 
Conversely, $\hat{P}_j = 1$ indicates a prediction of an entangled qubit pair, 
thus assigning it to the positive class.

\begin{figure}[tp!]
    \centering
    \includegraphics[width=0.9\linewidth]{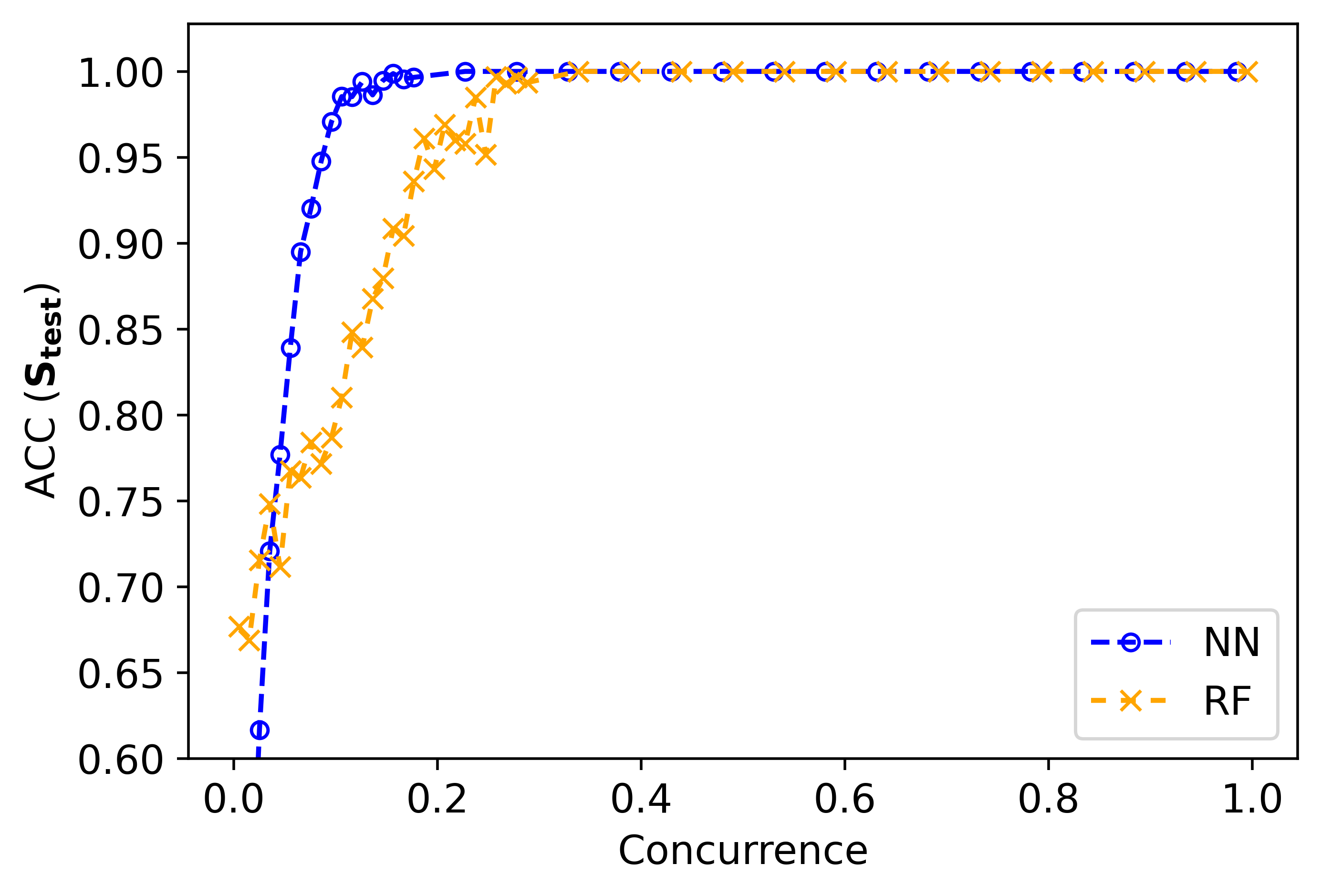}
    \caption{RF1 and NN model accuracy for the input states (from the test set $\Stest$) with different concurrence value.}
    \label{fig:acc_vs_conc}
\end{figure}

Let us analyze first the RF1 model. The accuracy of this model evaluated on $\Stest$ is
$\sim\!0.88$ (see Tab.~\ref{tab:rf_results}).
This value describes to overall performance of the model regardless of 
the concurrence value of the input samples.
To gain a deeper understanding of its error characteristics, we present Fig.~\ref{fig:acc_vs_conc}, which illustrates the relationship between the accuracy of the RF1 model, Eq.~(\ref{eq:rf_acc}), and the concurrence of the input sample.
The concurrence value depicted on the $x$-axis of Fig.~\ref{fig:acc_vs_conc} represents the ground truth value, calculated in accordance with Eq.~(\ref{eq:concurrence}).

In Fig.~\ref{fig:acc_vs_conc} it is evident that the RF1 model accuracy is at its lowest when the $C$ value approaches $0$, indicating minimal entanglement among the qubits. Consequently, the model is more likely to misclassify such inputs as separable cases. As the $C$ value rises, the model's accuracy correspondingly improves. It achieves an accuracy of $100\%$ when the concurrence of the input samples reaches $C \gtrsim 0.3$. We obtained similar findings for the RF2 and RF3 models. This observation is in accord with the t-SNE analysis of the data, illustrated in Fig.~\ref{fig:tSNE} in Appendix~\ref{apx:tsne}, which shows that entangled samples with elevated concurrence values are more prone to being well separated and, therefore, easily distinguishable 
from separable qubits.
On the other hand, if we take into account just separable (negative) samples from the $\Stest$,
i.e. those with $C(\rho) < \tau$, the accuracy of their correct classification is $\sim 0.86$. The misclassified negative samples might be confused with entangled samples of low concurrence value.
\begingroup
\setlength{\tabcolsep}{5pt}
\renewcommand{\arraystretch}{1.6}
\begin{table}[tp!]
    \centering    
    \begin{tabular}{c|cccc}
       model  & accuracy  & precision  & recall \\
       \hline
       RF1 & $0.87930$ & $0.87706$ & $0.90026$ \\
       RF2 & $0.86854$ & $0.88570$ & $0.86562$ \\
       RF3 & $0.85837$ & $0.86027$ & $0.87739$ \\
       MLP & $0.92212$ & $0.92794$ & $0.92608$ \\
    \end{tabular}
    \caption{Evaluation scores in entanglement classification for the RF classifiers and NN regressor on the $\Stest$ dataset.} 
    \label{tab:rf_results}
\end{table}
\endgroup

For the NN model, the prediction error metrics are defined as
\begin{equation}
    \rmse(\Stest) = \mse(\Stest)^{1/2}\,.
\end{equation}
In order to compare the NN model with the RF models, we define also accuracy for the NN regressor as
\begin{equation}
    \acc_\mathrm{NN}(\Stest) = 
    \Er_j\!\left[1-\left|\Theta(\tau - C_j)\! - \!\Theta(\tau_\mathrm{NN}-\hat{C}_j)\right|\right],
\end{equation}
where we explicitly count the number of correct classifications (instances where NN predictions match the data classes). The Heaviside function $\Theta(\cdot)$ compares output (or true) $C$ value with the assumed threshold. The NN classification threshold $\tau = 10^{-6}$ splits the predictions into separable (with $C_i\leq\tau$) and entangled ($C_i>\tau$) classes. The threshold $\tau_\mathrm{NN}=0.03$ was chosen to maximize the NN model accuracy by maximizing the area under the so-called precision-recall curve~\cite{Raghavan1989}. 
The resulting NN predictor performance, presented in Fig.~\ref{fig:acc_vs_conc}, is significantly better than RF over different $C$ values.

The overall comparison of the models is shown in Tab.~\ref{tab:rf_results}.
For the sake of completeness we also define the models' precision, which measures the
proportion of correctly identified entangled instances (true positives) out of 
all instances that were predicted to be entangled
\begin{equation}
    \prec\left(\Stest \right) = 
    \left. \Er_j\left[ P_j\, \hat{P_j} \right] \middle/ \Er_j \left[ \hat{P}_j \right] \right.\,,
\end{equation}
as well as recall, expressing the proportion of actual entangled instances that were correctly identified by the classifier
\begin{equation}
    \rec\left(\Stest \right) = 
    \left. \Er_j\left[ P_j\, \hat{P_j} \right] \middle/ \Er_j \left[ P_j \right] \right.\,.
\end{equation}
Notably, the NN model demonstrates superior performance compared to all of the RF models across the three metrics evaluated. 
This result is expected as deep NNs have proven to be more flexible and effective in entanglement recognition~\cite{Krawczyk2024,Pawlowski2024,Taghadomi2024,Chen2022,Asif2023,urena2024}, than classical,
non-neural ML algorithms~\cite{Goes2024,Lu2018}.
In the following Sections, we explore the underlying reasons. Among the RF classifiers, RF1, which comprises DTs with minimal regularization, appears as the most accurate one. Additional regularization of the DT sizes (present in RF2 and RF3, cf. Table~\ref{tab:rf_models} in Appendix~\ref{apx:rf}) leads to a decrease in the overall accuracy of the ensemble.

\subsection{Measurement importance}

We now direct our attention to the interpretability of the applied models. Here we assess the significance of the measurements that constitute our dataset.
We employ a variety of methods to assess feature importance, enabling a comparative analysis of data handling between ensemble and NN models.
Initially, we examine the MDI measure, defined in Appendix~\ref{apx:rf}, that is uniquely applicable to models derived from DTs. Consequently, it is not suitable for comparing ensemble models with NNs. Hence, we proceed to assess the significance of measurements through their \textit{perturbations}, facilitating a direct comparison between the two model types employed. Ultimately, we implement a more advanced approach predicated on \textit{Shapley values} (SVs).

\subsubsection{Mean Decrease in Impurity}

One of the key advantages of RF models is their ability to assess feature importance, providing insights into the contributions of different variables in predicting the target outcome. In an RF, feature importance is typically determined by measuring the impact of each feature on the model's predictive accuracy, e.g.~using Mean Decrease in Impurity (MDI)~\cite{Breiman2001} (see Appendix for details).

\begin{figure}[t]
    \centering
    \includegraphics[width=.96\columnwidth]{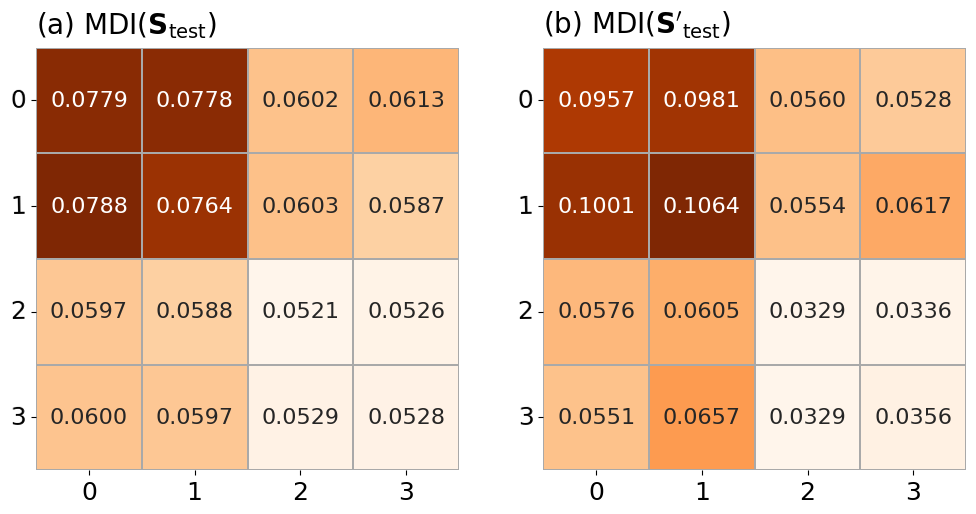}
    \caption{Mean Decrease in Impurity (MDI) of RF1 model trained and tested (a) using the samples with arbitrary values of concurrence, (b) using samples with concurrence $C < \tau$ (separable), and $C > 0.99$ (strongly entangled).}
    \label{fig:rf_feat_imp-heatmap}
\end{figure}
Fig.~\ref{fig:rf_feat_imp-heatmap}(a) shows MDI values obtained for the RF1 model, which has been trained utilizing the dataset $\Strain$. 
Within the heat map, each location, specified by indices $i$ and $j$ 
($i, j = 0, 1, 2, 3$),
corresponds to the $m_{ij}$ measurement, defined in Eq.~(\ref{Eq:measurements}). 
The data demonstrates that individual features exhibit varying levels of importance (measurements that are most crucial for sample classification possess the highest MDI values). Quite surprisingly, these measurements are situated in block $\bMA$, which only contains information about the diagonal elements of the density matrix
in a separable basis. Conversely, the lowest MDI values are found in block $\bMD$, which contains information about the off-diagonal elements which 
are critical for entanglement (non-local coherences). 
Blocks $\bMB$ and $\bMC$ encompass measurements of moderate significance.
It is relevant to note that the differences between the feature significance within the different blocks are moderate, and all features are essential to achieve accurate classification. 

The 
seemingly disproportionate importance of blocks that on their own do not contain information about entanglement (blocks $\bMA$, $\bMB$, and $\bMC$)
is due to two factors. Firstly, it is impossible to determine the non-local coherences from the measurements of block 
$\bMD$, without the knowledge 
of the other off-diagonal elements. This in turn can be found from the measurements of blocks $\bMB$ and $\bMC$ only when the occupations are known,
and occupations are directly measured via block $\bMA$.
Secondly, non-zero non-local coherences are only necessary for entanglement, while the actual entanglement 
strongly depends on the interplay between all elements of the density matrix. 

To determine if the importance of the different blocks is not a numerical peculiarity, we provide a second set of data where the RF1 model is trained 
on a subset of the training dataset, $\Strain' \subset \Strain$, which includes only states that are either separable ($C < \tau$) or strongly entangled ($C > 0.99$). The model is tested on an analogous subset of the test dataset, 
$\Stest' \subset \Stest$.
RF1 can distinguish between separable and
strongly entangled qubit pairs with a $100\%$ accuracy.
Fig.~\ref{fig:rf_feat_imp-heatmap}(b) shows the MDI values for this situation. 
We note that the relative importance of the different blocks remains unchanged,
but the features became significantly more prominent, so while the importance of blocks $\bMB$ and $\bMC$ remains similar,
the importance of measurements in $\bMA$ block increase,
while measurements in $\bMD$ have their significance further reduced.
This is highly surprising, but it does demonstrate that the hierarchy of feature importance is retained even if the analysis is performed on datasets
which are chosen to amplify the differences between separable and entangled states and is thus not a numerical peculiarity. 

\subsubsection{Measurement importance via perturbations}
\label{SSSec:perturb}

Since RF and NN models are fundamentally different approaches, our main goal is to determine how the NN predictor works compared to ensemble RF models.
To this end, we evaluate the importance of features of both models in the same way. The employed methodology for testing the importance of features (measurements) is presented in Fig.~\ref{fig:importance_sheme}(a).
For subsequent states from the test set $\Stest$, we perturb a given measurement by adding uniform noise $N_\sigma$ with interval $[-\sigma, \sigma]$ and then study how the prediction of the concurrence $C$ is perturbed. 
This way we describe the measurement importance in concurrence prediction. 

\begin{figure}[t]
    \centering
    \includegraphics[width=0.96\linewidth]{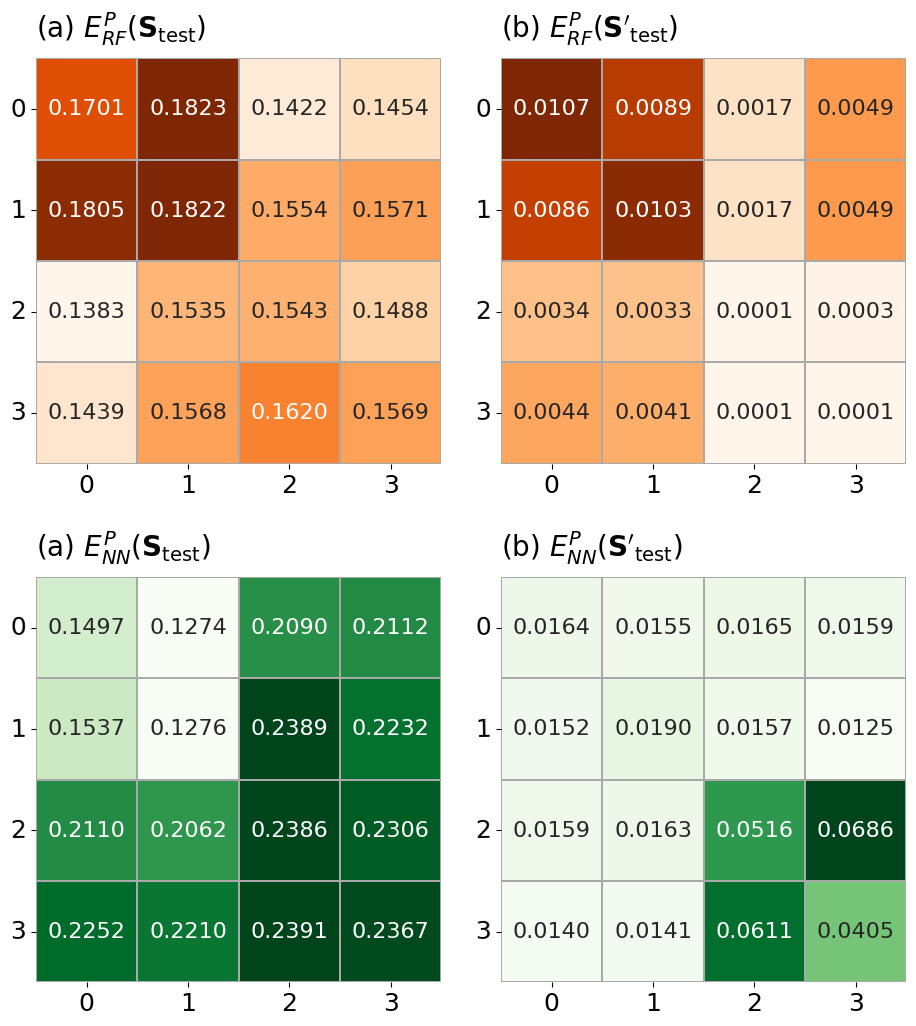}
    \caption{
    Perturbation measure $E^p$ for random noise applied to a single measurement outcome with $\sigma=0.05$ for (a) RF1 model trained using $\Strain$, (b) RF1 model trained using $\Strain'$ (i.e. samples with concurrence $\calc < 10^{-6}$ and $\calc > 0.99$); same for NN model trained using (c) $\Strain$, and (d) $\Strain'$.}
    \label{fig:rf_noise_test}
\end{figure}
Fig.~\ref{fig:rf_noise_test} summarizes the prediction fidelity when a single measurement characterized by indices $i,j=0,1,2,3$ is disturbed.
The RF classifier used is given by
\begin{equation}
    E^\mathrm{P}_\mathrm{RF}(\sigma, i) = 
    \left|\mathrm{ACC}_\mathrm{RF}(\Stest) - 
    \mathrm{ACC}_\mathrm{RF}(\tilde{\bm S}_\mathrm{test}(\sigma, i)) \right|\,,
\end{equation}
where $\tilde{\bm S}_\mathrm{test}$ is the test dataset modified by the perturbation in such a way that it only affects the measurement $m_{i,j}$.
It is perturbed by a random value from the
interval $[-\sigma, \sigma]$, as described above. 
Analogously, the NN classifier is defined as
\begin{equation}\label{eq:gtmeasure}
    E^\mathrm{P}_\mathrm{NN}(\sigma, i) = 
    \mathrm{E}_{jk}\!\left[ \left| \hat{C}_{ijk}-C_j \right| \right],
\end{equation}
where $C_j$ is the ground truth concurrence for the $\rho_j\in{}\Stest$ state. These measures estimate, how important a given feature (measurement) is for the prediction of the exact value of the state concurrence.

The data shows that the RF and NN models learn to predict their targets in very different way. 
The RF classifier shadows the counter-intuitive behavior of the feature importance classified by MDI, so it is most vulnerable to perturbations
of measurements in the $\bMA$ block, as seen in Fig.~\ref{fig:rf_noise_test}(a). There is a lesser difference between the effect of perturbations between the 
remaining three blocks. Results obtained on the datasets limited to separable and highly entangled states, Fig.~\ref{fig:rf_noise_test}(b), recapture all of the
MDI features, thus proving that feature importance classified by MDI and the effect of perturbations produce the same result. 

Contrarily, the NN results for both the full dataset, Fig.~\ref{fig:rf_noise_test}(c), and the restricted one, Fig.~\ref{fig:rf_noise_test}(d), show fully intuitive behavior.
The block containing information about the non-local coherences ($\bMD$) is the most important, 
closely followed by blocks with information about other off-diagonal elements ($\bMB$ and $\bMC$), while the block that only provides statistical information
($\bMA$) is comparatively irrelevant. 

These results highlight that the two types of models work very differently in entanglement quantification for tomography. The NN model is more disturbed by the perturbations
of the measurements which directly contain information about the more relevant elements of the density matrix. The RF model obtains it's results in a more
convoluted way and the features that in themselves contain no information about entanglement are most relevant through their interplay with the information
contained in other measurements. 

Incidentally, such a discrepancy in feature importance does not occur when the two 
techniques are used to quantify the concurrence directly from the elements of the density matrix.
In such a case, both RF and NN models indicate that the non-local coherences are of highest importance, followed by the other off-diagonal elements, and then by occupations. 
This means that the difference between the models stems from the way that 
information about the density matrix is inferred from the measurement outcomes.

In addition, in Appendix \ref{Sec:pert_sigma} we compare the average performance of RF and NN models under a single measurement perturbation. It is shown, that NN model dominates at lower noise levels. However, as the noise amplitude increases, the the accuracy of NN model decreases faster than the one of RF. As a result, the RF model outperforms the NN at higher noise level (see Fig.~\ref{fig:pert_measure_sigma}).
RF models combine the predictions of many decision trees trained on bootstrap‑resampled subsets of data. Because each tree sees a different slice of the input features, their averaged vote smooths out individual errors, making the overall model noticeably sturdier against noisy features and labels.

\subsubsection{Shapley additive explanations}


In the previous section we have shown that RFs and NNs make their predictions in a way which is sensitive to 
different measurements at the input.
Thus, we compute the so-called Shapley values~\cite{shapley1953} to explain how the models learn and predict.
SVs, originally developed in cooperative game theory, provide a fair way to distribute the total contribution of all players (or features) in a system by considering all possible coalitions. 
In the context of ML, SVs offer a way to quantify the contribution of each feature to the model’s prediction~\cite{Lundberg2017}. 
Importantly, SVs form a model-agnostic method, 
meaning they can be applied to any ML model without requiring specific modifications. Detailed definitions of SVs and Shapley interaction values (SIVs) are described in Appendix~\ref{apx:shap}. 
To compute SVs and SIVs, we employ the SHAP (Shapley additive explanations) library~\cite{Lundberg2017}, which provides a unified framework for interpreting ML models based on SVs.

\begin{figure}[tp]
    \centering
    \includegraphics[width=0.96\linewidth]{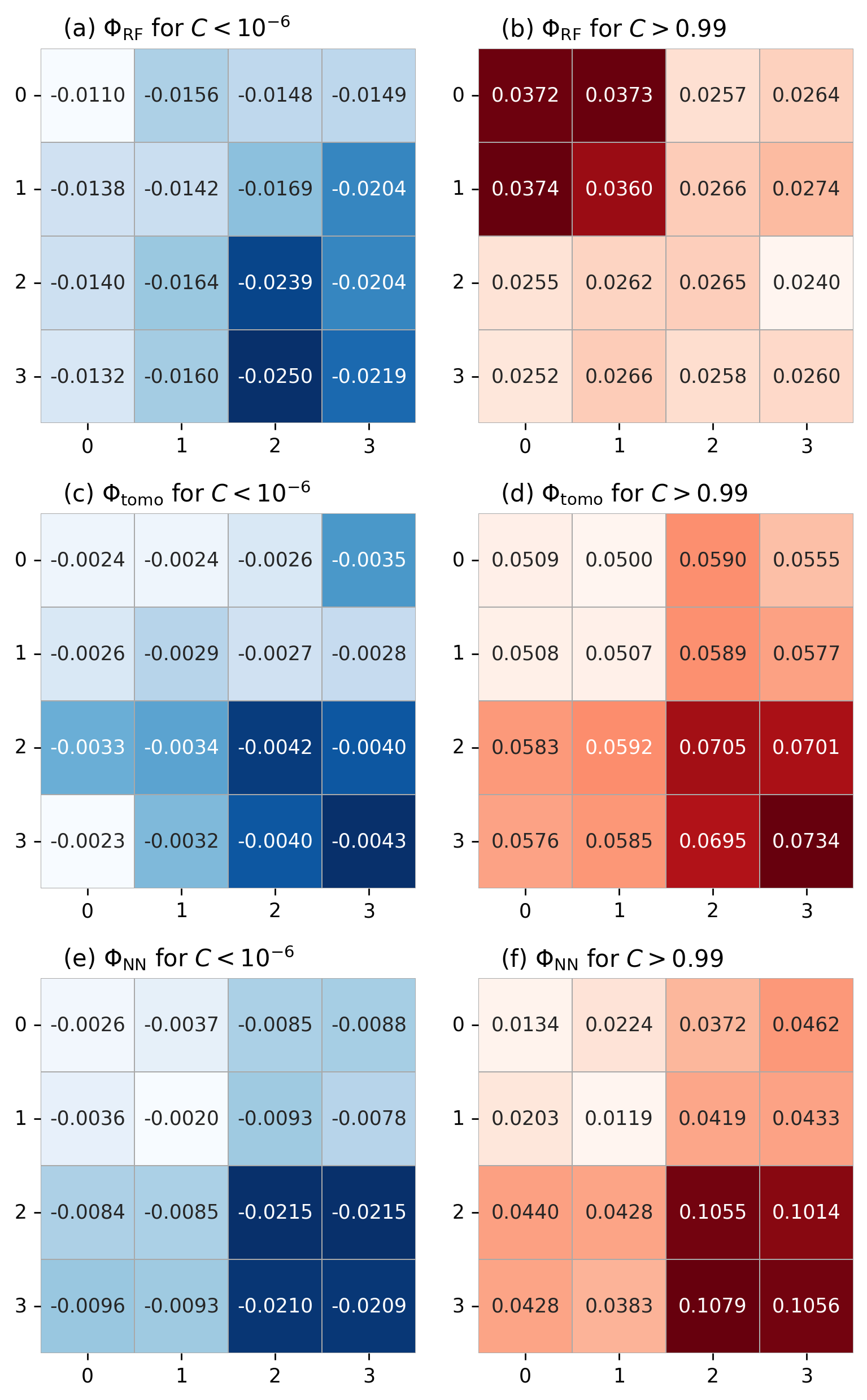}
    \caption{Shapley values $\Phi_i$ calculated using (a,c,e) samples with concurrence $C < \tau$ (separable), and (b,d,f) $C > 0.99$ (strongly entangled), for (a,b) RF1 model, (c,d) states reconstructed via quantum tomography, and (e,f) NN predictor.
    }
    \label{fig:shap}
\end{figure}

Let us first analyze the SVs of the RF model.
Since SVs provide an explanation for a single sample, 
they allow us to analyze how the RF model classifies samples with concurrence in a given range of values. 
Figs \ref{fig:shap}(a,b) show SVs averaged over a subset of samples from the testing dataset (a) with concurrence below the threshold $\tau$, i.e. for separable qubits, and (b) with concurrence above $0.99$.
In our sign notation, the positive mean Shapley value means, that the corresponding feature contributes to increasing the likelihood of class $1$ (entangled). Oppositely,
when a Shapley value is negative it drives the prediction towards class $0$ (separable).
Thus, it is obvious that the model treats the two groups in a very different manner. While the most important features contributing to correct classification of separable samples are those in segment $\bMD$, the strongly entangled samples are classified based on the measurement values form set $\bMA$. 

In addition, the contribution of feature interactions -- cases where features jointly influence the prediction in a way that goes beyond their individual effects -- is studied in Appendix \ref{apx:shap_int_rf} using the Shapley interaction values (SIVs). 
In case of separable states, it is shown that diagonal SIVs pushes the model's prediction towards the opposite class (entagled). The contribution leading to the correct prediction  is originating from the feature interactions. Among them the interactions of the measurements ($m_{00}$, $m_{11}$) and ($m_{01}$, $m_{10}$) are dominant.
Conversely, if we take into account just the strongly entangled states, the model prediction is based mainly on the single measurement values.

Before we start the analysis of SVs for the NN model, let us look at Figs.~\ref{fig:shap}(c,d) which show SVs for the concurrence $C$ calculated from QST, i.e., 
where the concurrence is calculated directly for a state reconstructed from measurements via quantum tomography analytically. Here, all the measurements are of similar importance, with the $\bMA$ block slightly less relevant than the others, which is intuitive from the point of view of quantum information theory.

SVs of the NN model presented in Figs.~\ref{fig:shap}(e,f) follow the same behavior as the QST-reconstructed results.
(cf. Fig.~\ref{fig:shap}(c,d)).
The $\bMD$ block has the most impact on $C$ prediction. Interestingly $\bMD$ increases $C$ predictions for separable states, see Fig.~\ref{fig:shap}(e), but decreases for entangled ones -- Fig.~\ref{fig:shap}(f).
$\bMB$ and $\bMC$ blocks are slightly weaker but still more important than block $\bMA$, which only contains information about the occupation of separable states.
Note that similar structure of measurement importance for NN was also observed with the perturbation measure $E^P$, as presented in Fig.~\ref{fig:rf_noise_test}(c).
We arrive at the intriguing result that, depending on the type of ML model (RF vs. NN), different measurement groups are relevant for the  $C$ prediction
which strongly suggests that the two types of machine learning models operate very differently when trying to evaluate the amount of entanglement 
present in a given state. 

\section{Conclusions\label{sec8}}

We have studied the performance of NN and RF models in the quantification of entanglement for two qubits from quantum tomography data. We found that NN networks, which are known to be more flexible and efficient in entanglement prediction from quantum states, perform better also when tomography measurement outcomes 
are given on input instead of density matrix elements.

More surprisingly, we found that the importance of different measures differs greatly between
the two classes of ML algorithms. NN feature importance follows the intuitive
behavior in which the most relevant are measurements that carry information about the
qubit coherences which appear e.g.~in Bell or X states, followed measurements that capture other coherences, and lastly the ones responsible for diagonal elements of the density matrix
(written in a separable basis). The same feature importance is exhibited when the concurrence
is calculated directly from QSR reconstruction (i.e. without ML), which we assessed using Shapley values. RF feature 
importance is fully contradictory: the most important measurements are the ones which directly
yield the diagonal elements and the ones that carry information about inter-qubit coherence
were found to be least important. This points to a vast difference in the way that
the two ML techniques acquire information from the tomography measurements. 
When the same analysis was performed for learning of the concurrence directly from
elements of the density matrix, such a discrepancy was not observed.

Neural networks have the ability to build efficient representations (intrinsic dimensionality reduction) using large data vectors, while classical ML models need less better quality and more universal features and cannot cope with more diffused information. Therefore, the RF model chose the measurements that are more relevant for the extraction of information
(also from other measurements), whereas the NN model behaves in accordance with the QST, being able to effectively predict concurrence from the full measurement set.

On the other hand, the RF models can outperform the networks in the presence of noise perturbing a single measurement, as demonstrated in Appendix \ref{Sec:pert_sigma}.
The higher robustness of RF model to the noise perturbation stems from the cooperation of a big number of classifiers relying on different features of the input vector.

In general, any model performs better when all measurements are reliable, but the different feature importance displayed by the models which translated into corresponding behavior in the presence of noise on the corresponding measurement, suggests that the choice of ML strategy is of vital importance when some measurements are likely to be less reliable. ML models can be used to test the universality and interdependence of measurements as an information resource for quantum state reconstruction.

\section*{Acknowledgments}

KR is funded within the
QuantERA II Programme that has received funding from the
EU H2020 research and innovation programme under Grant
Agreement No. 101017733, and with funding organization
MEYS.
MK and JP acknowledge support from National Science Centre, Poland, under grant no. 2021/43/D/ST3/01989.
We gratefully acknowledge Polish high-performance computing infrastructure PLGrid (HPC Centers: ACK Cyfronet AGH) for providing computer facilities and support within computational grant no. PLG/2024/017284.
This work was supported by the research programme of the Strategy AV21 AI: Artificial Intelligence for Science and Society.

\appendix

\section{Dataset inspection}

In order to have a more comprehensive picture of the studied datasets, we 
analyzed the data using standard unsupervised ML techniques for feature vector dimensionality reduction: PCA and the more recent t-SNE.

\subsection{Principal Component Analysis}\label{apx:pca}

At first, we examined the training dataset using PCA method~\cite{Jolliffe2002} --
an unsupervised dimensionality reduction technique used to reduce the dimensionality of complex datasets 
while retaining their most important characteristics.
PCA finds a linear transformation of the original features into a set of new variables, called principal components.
The principal components are ordered by their {\em explained variance ratio}, which defined as the
amount of variance they capture in the data.
Explained variance ratio helps us determine how many principal components to retain for analysis, as components with higher explained variance ratios contain more information about the dataset's structure.

\begin{figure}[tp]
	\centering
	\includegraphics[width=.96\columnwidth]{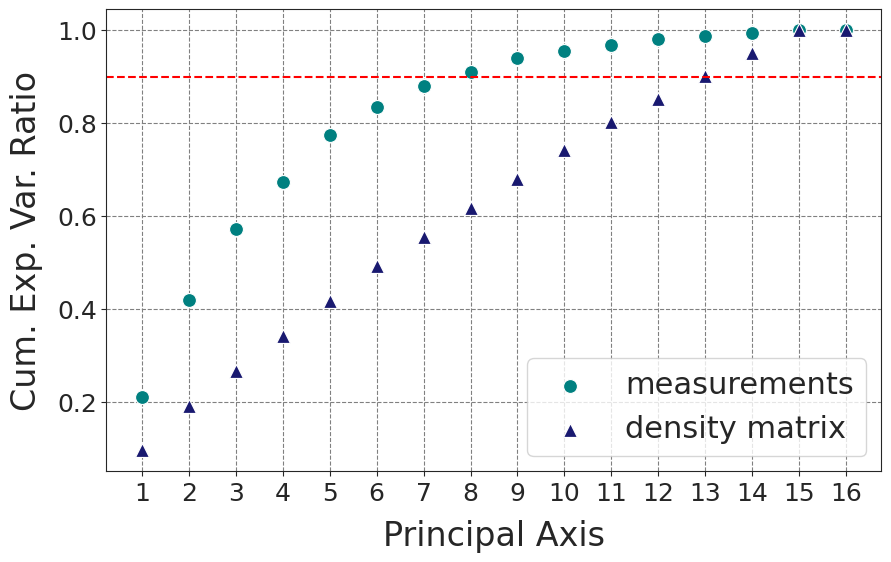}
	\caption{Cumulative explained variance ratio of PCA performed on datasets vectors of measurements (dots) and on components of density matrices (triangles).
		The red dashed lines show the boundary of $90\%$ of preserved information.}
	\label{fig:pca}
\end{figure}

Our goal is to estimate how many components of the feature vectors do we need to preserve majority of the information in the dataset. To this end we performed PCA on the training dataset and evaluated the explained variance ratios of each component, $r_i$, where $i = 1, 2, \dots, 16$.
For comparison, we performed PCA on both training datasets consisted of measurement outcomes as well as on the components of the density matrices. 
Fig.~\ref{fig:pca} compares cumulative explained variance ratio,
$R_j$, of PCA performed on both datasets, defined as
\begin{equation}
	R_j = \sum_{i=1}^{j} r_i\,, 
	\quad \text{where} \quad
	j = 1, 2, \dots, 16\,.
	\label{eq:cum_exp_var_rat}
\end{equation}
By definition the total sum of explained variance ratios equals $1$, thus $R_{16} = 1$.
Obviously, both dependencies strongly differ. 
In case of the dataset consisting of the density matrix elements, we need $13$ principal components to reach the boundary of $0.9$, which corresponds to $90\%$ of preserved information. On the other hand, when we deal with measurements outcomes, we need just first $8$ PCA components to overcome this threshold.
This observation speaks in favour of the possibility of quantum tomography based on a limited number of measurements.
In addition, we notice that in case of both datasets we reach cumulative explained variance ration equal to $1$ for $15$ principal components. 
This stems from the fact that the density matrix components are not independent. Namely, the trace equals $1$, which allows one to calculate one diagonal component using the others. 


\subsection{t-Distributed Stochastic Neighbor Embedding}\label{apx:tsne}

Second, we employed the t-Distributed Stochastic Neighbor Embedding (t-SNE) method~\cite{hinton2003sne}, which is a
dimensionality reduction technique particularly well suited for visualizing high-dimensional data by mapping them into a lower-dimensional space. Importantly, it captures the local structure of the data while preserving global relationships in the dataset.

\begin{figure}[tb]
	\centering
	\includegraphics[width=0.9\columnwidth]{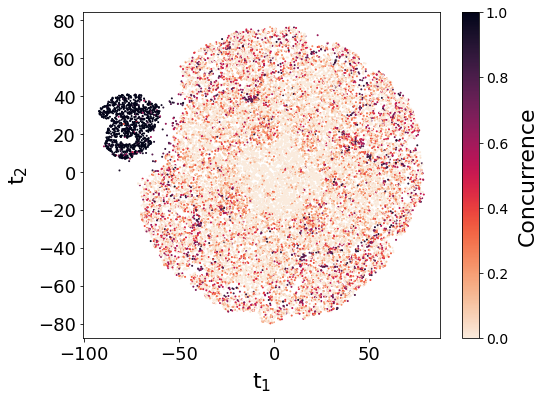}
	\caption{Measurement vectors projected into 2D space using t-SNE method. Note that both dimensions $t_{1,2}$ defining embedding space do not have strict interpretation.}
	\label{fig:tSNE}
\end{figure}
Fig.~\ref{fig:tSNE} shows the projection of measurement vectors in a 2D embedding space.
Analyzing the t-SNE embedded space one can see that the entangled states seem to cluster nicely, but is worth noting that they also occur largely between separable states (as black dots in the bright cluster). Thus, the detection of entanglement based on this representation is worse than the ones obtained using RF or NN predictors.

\section{Random Forest details}\label{apx:rf}

Each DT in a RF constitutes a tree graph structure in which each node signifies a decision rule, while each leaf node corresponds to a specific class.
The \textit{purity} of the $k$-th node is specified by the Gini impurity~\cite{Breiman2001}, $G_k$, defined as
\begin{eqnarray}
    G_k = 1 - \sum_{i=1}^N p_{i, k}^2\,,
\end{eqnarray}
where $N$ is the number of classes, and $p_{i, k}$ is the proportion of samples of class $i$ in the node $k$.
The Mean Decrease in Impurity~\cite{Breiman2001,scornet2023trees} can be calculated from the training samples and serves as a metric within DT-based models to ascertain feature importance. It quantifies the extent to which a feature contributes to the reduction of Gini impurity in the dataset when it is utilized for data splitting at a node within the tree.

For this purpose, the {\tt RandomForestClassifier} from the Scikit-learn library~\cite{scikit-learn} was employed. We applied a grid search algorithm alongside cross-validation to fine-tune the model's hyperparameters. The most effective RF hyperparameters are presented in Table~\ref{tab:rf_models}, where {\tt n\_estimators} denotes the number of DTs within the ensemble. During the search for optimal splits, only a random subset of features is considered, with its size determined by {\tt max\_features}. The boolean parameter {\tt bootstrap} indicates whether bootstrap samples are utilized in tree construction. Additionally, {\tt max\_depth} represents the maximum allowable tree depth, and {\tt min\_samples\_leaf} defines the minimum number of samples required for a leaf node. Should the aforementioned parameters be configured to {\tt None}, this constraint is not factored into the regularization process.
\begingroup
\setlength{\tabcolsep}{10pt}
\renewcommand{\arraystretch}{1.6}
\begin{table}[tp!]
    \centering
    \begin{tabular}{c|ccc}
        model & RF1 & RF2 & RF3 \\
        \hline
        {\tt n\_estimators}  & $1000$ & $1000$ & $1000$ \\
        {\tt max\_features}  & $4$ & $16$ & $4$ \\
        {\tt max\_depth} & {\tt None} & $20$ & {\tt None} \\
        {\tt min\_samples\_leaf} & {\tt None} & {\tt None} & $100$\\
        {\tt bootstrap} & {\tt True} & {\tt True} & {\tt True}\\
    \end{tabular}
    \caption{Details of the hyperparameters of the most successful Random Forest models. The explanations of these hyperparameters can be found within the main text.}
    \label{tab:rf_models}
\end{table}
\endgroup

\section{Definition of Shapley values and Shapley interaction values}\label{apx:shap}
The Shapley value, SV for a feature $i$ in a model $f$ is defined as
\begin{equation}
    \phi_i = \sum_{S \subseteq N \setminus \{i\}} w_1(S, N)\, \Delta f(S, i),
\end{equation}
where $w_1(S, N) = |S|!(|N|-|S|-1)! / |N|!$ is a weighting factor 
with $N$ being the set of all features, and 
$S$ representing a subset of features excluding $i$.
\begin{equation}
    \Delta f(S, i) = f(S \cup \{i\}) - f(S)
\end{equation}
represents the marginal contribution of feature $i$.
This definition ensures that contributions are fairly allocated based on marginal contributions across all possible subsets.

For binary classification problems, SVs can be computed separately for each class. The SVs for class 0 and class 1 are equal in magnitude but differ in sign,
$\phi_i(1) = -\phi_i(0)$.
This sign of SV $\phi_i(c)$ reflects the contribution of a feature towards increasing or decreasing the probability of a given class, $c$. A positive SV for class $1$ indicates that the feature increases the probability of predicting class $1$, while a negative value suggests it supports class $0$ instead.

Importantly, SVs are computed for individual samples, providing a local explanation of how a particular instance is classified. To obtain a global understanding of feature importance across the dataset, we compute the \textit{global} SVs over all instances in a selected set of samples, 
${\bm X} \subset \Stest$,
\begin{equation}
    \Phi_i = E_k\left[ \phi_i^{(k)} \right],
\end{equation}
where $\phi_i^{(k)}$ represents the SV of feature $i$ for sample $X_k \in {\bm X}$.

While SVs explain individual feature contributions, they do not capture interactions between features. 
The Shapley interaction value, SIV extends the concept by quantifying the pairwise interactions between features in a model. 
It measures how much the joint presence of two features contributes to the prediction beyond their individual effects.

Similarly, to single value SVs, 
a SIV for two features $i$ and $j$ in a model $f$ is defined as
\begin{equation}
    \phi_{i,j} = \sum_{S \subseteq N \setminus \{i,j\}} w_2(S, N)\, \Delta f(S, i, j),
\end{equation}
where the weighting factor for subset size is
$w_2(S, N) = |S|!(|N|-|S|-2)! / 2(|N|!)$ is, and 
\begin{equation}
    \begin{split}
    \Delta f(S, i, j) = &f(S \cup \{i,j\})\, - \\
    & f(S \cup \{i\}) - f(S \cup \{j\}) + f(S)
    \end{split}
\end{equation}
quantifies the additional contribution of both features appearing together.
This formulation ensures that the interaction term captures the additional contribution of both features appearing together, beyond what would be expected from their individual SVs. SIVs are particularly useful for understanding dependencies between features and identifying synergistic or redundant effects in complex models. 

The \textit{global} Shapley interaction values, SIVs are defined as
\begin{equation}
    \Phi_{i, j} = E_k\left[ \phi_{i, j}^{(k)} \right]\,,
\end{equation}
where $\phi_{i, j}^{(k)}$ is the SIV of
features $i$ and $j$ for sample $X_k \in {\bm X}$.

SIVs are closely related to standard SVs. The total contribution of a feature $i$ can be decomposed into its individual SV and its interactions with other features
\begin{equation}
\phi_i = \phi_i^{\text{ind}} + \sum_{j \neq i} \phi_{i,j}
\end{equation}
where $\phi_i^{\text{ind}}$ represents the independent contribution of feature $i$, and
the second term accounts for the total interaction effects with other features.
This decomposition highlights that the SV of a feature includes both its standalone effect and all pairwise interactions with other features. If there are no significant interactions, then $\phi_{i,j} \approx 0$, reducing the model to an additive form where each feature contributes independently.

In this study, we have calculated SVs and SIVs for the RF classifier using the SHAP’s \texttt{TreeExplainer} class. The \texttt{TreeExplainer} is specifically optimized for tree-based models, leveraging their structure to efficiently compute SVs and interactions~\cite{Lundberg2017}.

\section{Shapley interaction values for RFs}\label{apx:shap_int_rf}

\begin{figure}[htp!]
    \centering
    \includegraphics[width=0.99\linewidth]{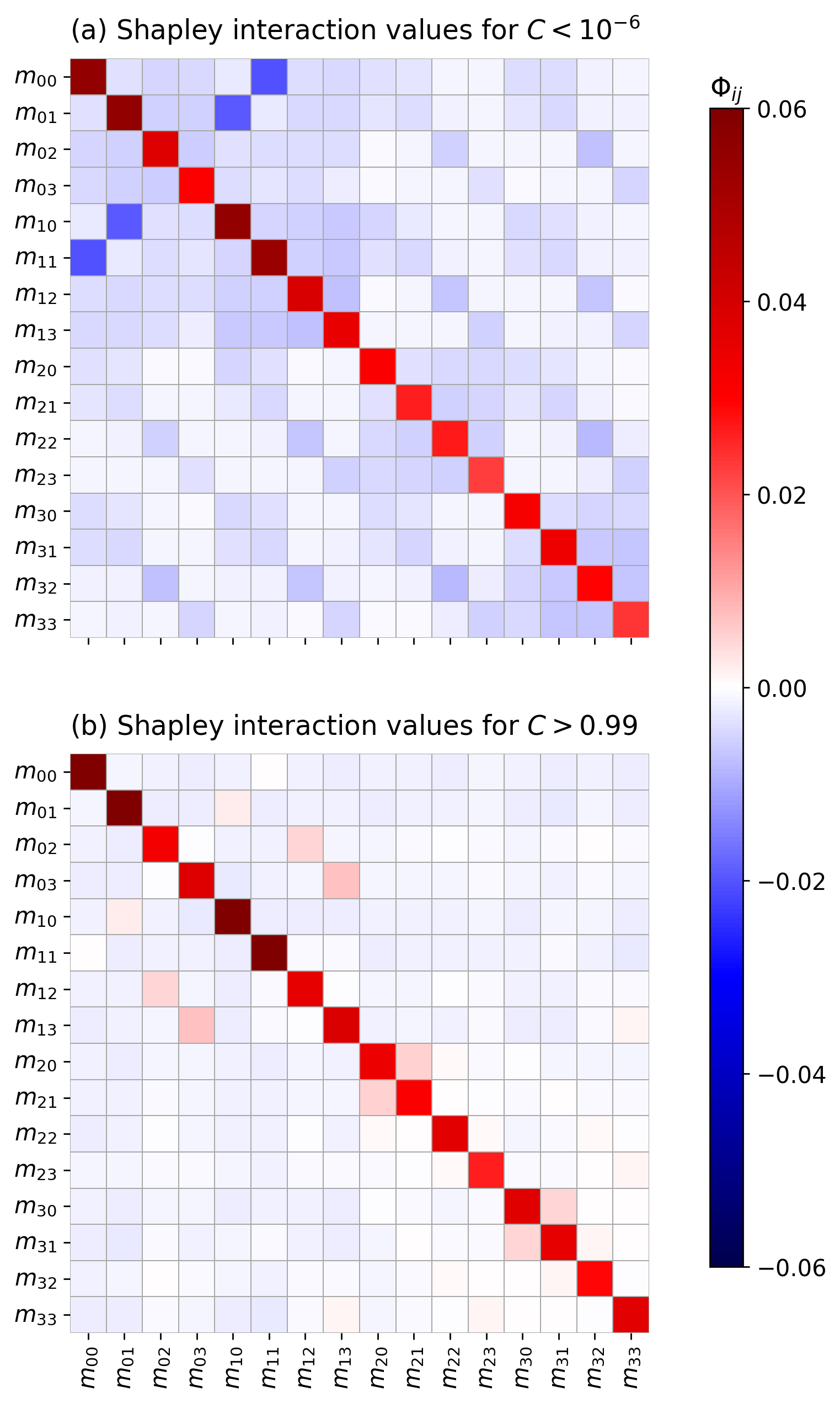}
    \caption{Global SIVs calculated for a simplified Random Forest model using a subset of $\Stest$ with 
    (a) $C < 10^{-6}$ (separable) and (b) $C > 0.99$ (strongly entangled).}
    \label{fig:shap_int_rf}
\end{figure}

To elucidate different approach of the RF classifier to separable and entangled samples, we extend our analyses to SIVs. The global SIVs for (a) separable, and (b) strongly entangled samples are shown in Fig.~\ref{fig:shap_int_rf}.
Since the computational costs of SIVs for a large model with $16$ features are enormous, we did the calculations
using a smaller RF model consisted of $100$ decision trees regularized by {\tt min\_samples\_leaf} $= 20$.
After training of the model using $\bS_{\rm train}$ dataset, it shows Shapley values similar to those calculated for the RF1 model. Thus, we believe that the SIVs computed for the simplified model might shed some light on the performance of the large RF models studied in this paper.

If a SIV in Fig.~\ref{fig:shap_int_rf} is positive, it drives the prediction of the model towards class $1$ (entangled),
while the negative SIVs increase the probability of predicting class $0$ (separable).
Thus, the first look on the case with strongly entangled samples reveals, that the classifier
classifies these samples based on the single features.
Among them the 4 measurements in the block $\bMA$ appears to be most important, 
which is in accord with our previous results for RF classifier.
In addition, one might observe minor contribution of interactions between features. 
All the other SIVs are close to zero.
On the other hand, in case of separable samples, the diagonal elements of the Shapley interaction matrix 
drives the predictions towards the opposite class (entangled). In this case, however,
the strong contribution of features interactions prevails and draw the prediction towards to correct result.
Among the interactions, the most important appears to be the interaction between measurements
($m_{00}$, $m_{11}$) and ($m_{01}$, $m_{10}$), which are also located in block $\bMA$.

The observations from Fig.~\ref{fig:shap}(a) and (b) shows that in case of the samples with $C > 0.99$ the most important features are in the $\bMA$ block, while for classification of the separable samples the most important measurements are located in $\bMD$ block. Fig. \ref{fig:shap_int_rf} shows that the structure of the diagonal SIV does not significantly changes for the separable and strongly correlated samples. The crucial effect have the negative contributions of the feature interactions, which compensate the strong positive effect of diagonal elements.
Moreover, this result suggests that the RF models tend to identify the separable samples not based on the single measurements but on the mutual correlations of the features.

\section{Perturbation measure}
\label{Sec:pert_sigma}

\begin{figure}[h!]
    \vspace{5mm}
    \centering
    \includegraphics[width=0.99\linewidth]{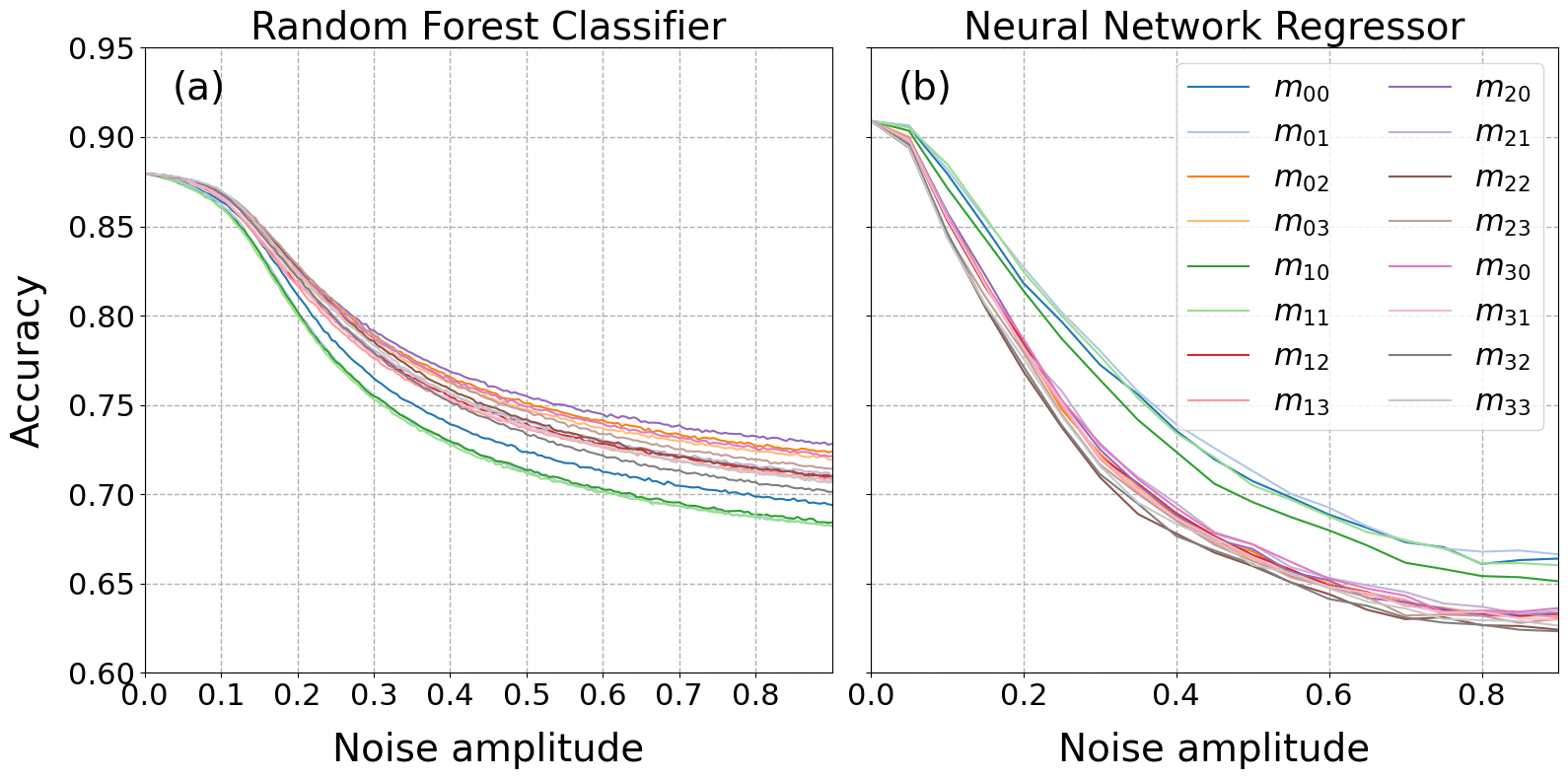}
    \caption{Perturbation measure for (a) random forest classifier, and 
    (b) neural network as a function of the noise amplitude $\sigma$.}
    \label{fig:pert_measure_sigma}
\end{figure}

In addition, we compare the performance of random forest and neural network for random noise applied to a single measurement outcome, as described in section \ref{SSSec:perturb}.
Fig.~\ref{fig:pert_measure_sigma} shows accuracy calculated for RF1 and NN models.
Although in case of zero noise, the accuracy of NN exceeds the one of RF model, 
in the presence of certain noise level the RF model might in average perform better.

The reason for this behavior is the large number of classifiers in the RF ensemble, which makes the method more robust against noise and errors. Because each tree in the RF ensemble classifies a sample in a different way, a random change of one feature is unlikely to affect prediction of every tree in the ensemble. Therefore, even stronger noise in the data does not significantly affect the overall performance of the model.
On the other hand, NN is a highly nonlinear model, which often relies not just on the single parameters, but also on their correlations. Thus, a systematic perturbation of one of the correlated features might significantly reduce the model accuracy.


\begin{thebibliography}{46}%
	\makeatletter
	\providecommand \@ifxundefined [1]{%
		\@ifx{#1\undefined}
	}%
	\providecommand \@ifnum [1]{%
		\ifnum #1\expandafter \@firstoftwo
		\else \expandafter \@secondoftwo
		\fi
	}%
	\providecommand \@ifx [1]{%
		\ifx #1\expandafter \@firstoftwo
		\else \expandafter \@secondoftwo
		\fi
	}%
	\providecommand \natexlab [1]{#1}%
	\providecommand \enquote  [1]{``#1''}%
	\providecommand \bibnamefont  [1]{#1}%
	\providecommand \bibfnamefont [1]{#1}%
	\providecommand \citenamefont [1]{#1}%
	\providecommand \href@noop [0]{\@secondoftwo}%
	\providecommand \href [0]{\begingroup \@sanitize@url \@href}%
	\providecommand \@href[1]{\@@startlink{#1}\@@href}%
	\providecommand \@@href[1]{\endgroup#1\@@endlink}%
	\providecommand \@sanitize@url [0]{\catcode `\\12\catcode `\$12\catcode
		`\&12\catcode `\#12\catcode `\^12\catcode `\_12\catcode `\%12\relax}%
	\providecommand \@@startlink[1]{}%
	\providecommand \@@endlink[0]{}%
	\providecommand \url  [0]{\begingroup\@sanitize@url \@url }%
	\providecommand \@url [1]{\endgroup\@href {#1}{\urlprefix }}%
	\providecommand \urlprefix  [0]{URL }%
	\providecommand \Eprint [0]{\href }%
	\providecommand \doibase [0]{https://doi.org/}%
	\providecommand \selectlanguage [0]{\@gobble}%
	\providecommand \bibinfo  [0]{\@secondoftwo}%
	\providecommand \bibfield  [0]{\@secondoftwo}%
	\providecommand \translation [1]{[#1]}%
	\providecommand \BibitemOpen [0]{}%
	\providecommand \bibitemStop [0]{}%
	\providecommand \bibitemNoStop [0]{.\EOS\space}%
	\providecommand \EOS [0]{\spacefactor3000\relax}%
	\providecommand \BibitemShut  [1]{\csname bibitem#1\endcsname}%
	\let\auto@bib@innerbib\@empty
	\bibitem [{\citenamefont {James}\ \emph {et~al.}(2001)\citenamefont {James},
		\citenamefont {Kwiat}, \citenamefont {Munro},\ and\ \citenamefont
		{White}}]{James2001}%
	\BibitemOpen
	\bibfield  {author} {\bibinfo {author} {\bibfnamefont {D.~F.~V.}\
			\bibnamefont {James}}, \bibinfo {author} {\bibfnamefont {P.~G.}\ \bibnamefont
			{Kwiat}}, \bibinfo {author} {\bibfnamefont {W.~J.}\ \bibnamefont {Munro}},\
		and\ \bibinfo {author} {\bibfnamefont {A.~G.}\ \bibnamefont {White}},\
	}\bibfield  {title} {\bibinfo {title} {Measurement of qubits},\ }\href
	{https://doi.org/10.1103/PhysRevA.64.052312} {\bibfield  {journal} {\bibinfo
			{journal} {Phys. Rev. A}\ }\textbf {\bibinfo {volume} {64}},\ \bibinfo
		{pages} {052312} (\bibinfo {year} {2001})}\BibitemShut {NoStop}%
	\bibitem [{\citenamefont {Torlai}\ \emph {et~al.}(2018)\citenamefont {Torlai},
		\citenamefont {Mazzola}, \citenamefont {Carrasquilla}, \citenamefont
		{Troyer}, \citenamefont {Melko},\ and\ \citenamefont {Carleo}}]{Torlai2018}%
	\BibitemOpen
	\bibfield  {author} {\bibinfo {author} {\bibfnamefont {G.}~\bibnamefont
			{Torlai}}, \bibinfo {author} {\bibfnamefont {G.}~\bibnamefont {Mazzola}},
		\bibinfo {author} {\bibfnamefont {J.}~\bibnamefont {Carrasquilla}}, \bibinfo
		{author} {\bibfnamefont {M.}~\bibnamefont {Troyer}}, \bibinfo {author}
		{\bibfnamefont {R.}~\bibnamefont {Melko}},\ and\ \bibinfo {author}
		{\bibfnamefont {G.}~\bibnamefont {Carleo}},\ }\bibfield  {title} {\bibinfo
		{title} {Neural-network quantum state tomography},\ }\href
	{https://doi.org/10.1038/s41567-018-0048-5} {\bibfield  {journal} {\bibinfo
			{journal} {Nature Physics}\ }\textbf {\bibinfo {volume} {14}},\ \bibinfo
		{pages} {447} (\bibinfo {year} {2018})}\BibitemShut {NoStop}%
	\bibitem [{\citenamefont {Xin}\ \emph {et~al.}(2019)\citenamefont {Xin},
		\citenamefont {Lu}, \citenamefont {Cao}, \citenamefont {Anikeeva},
		\citenamefont {Lu}, \citenamefont {Li}, \citenamefont {Long},\ and\
		\citenamefont {Zeng}}]{Xin2019}%
	\BibitemOpen
	\bibfield  {author} {\bibinfo {author} {\bibfnamefont {T.}~\bibnamefont
			{Xin}}, \bibinfo {author} {\bibfnamefont {S.}~\bibnamefont {Lu}}, \bibinfo
		{author} {\bibfnamefont {N.}~\bibnamefont {Cao}}, \bibinfo {author}
		{\bibfnamefont {G.}~\bibnamefont {Anikeeva}}, \bibinfo {author}
		{\bibfnamefont {D.}~\bibnamefont {Lu}}, \bibinfo {author} {\bibfnamefont
			{J.}~\bibnamefont {Li}}, \bibinfo {author} {\bibfnamefont {G.}~\bibnamefont
			{Long}},\ and\ \bibinfo {author} {\bibfnamefont {B.}~\bibnamefont {Zeng}},\
	}\bibfield  {title} {\bibinfo {title} {Local-measurement-based quantum state
			tomography via neural networks},\ }\href
	{https://doi.org/10.1038/s41534-019-0222-3} {\bibfield  {journal} {\bibinfo
			{journal} {npj Quantum Information}\ }\textbf {\bibinfo {volume} {5}},\
		\bibinfo {pages} {109} (\bibinfo {year} {2019})}\BibitemShut {NoStop}%
	\bibitem [{\citenamefont {Melkani}\ \emph {et~al.}(2020)\citenamefont
		{Melkani}, \citenamefont {Gneiting},\ and\ \citenamefont
		{Nori}}]{Melkani2020}%
	\BibitemOpen
	\bibfield  {author} {\bibinfo {author} {\bibfnamefont {A.}~\bibnamefont
			{Melkani}}, \bibinfo {author} {\bibfnamefont {C.}~\bibnamefont {Gneiting}},\
		and\ \bibinfo {author} {\bibfnamefont {F.}~\bibnamefont {Nori}},\ }\bibfield
	{title} {\bibinfo {title} {Eigenstate extraction with neural-network
			tomography},\ }\href {https://doi.org/10.1103/PhysRevA.102.022412} {\bibfield
		{journal} {\bibinfo  {journal} {Phys. Rev. A}\ }\textbf {\bibinfo {volume}
			{102}},\ \bibinfo {pages} {022412} (\bibinfo {year} {2020})}\BibitemShut
	{NoStop}%
	\bibitem [{\citenamefont {Ahmed}\ \emph {et~al.}(2021)\citenamefont {Ahmed},
		\citenamefont {S\'anchez Mu\~noz}, \citenamefont {Nori},\ and\ \citenamefont
		{Kockum}}]{Shahnawaz2021}%
	\BibitemOpen
	\bibfield  {author} {\bibinfo {author} {\bibfnamefont {S.}~\bibnamefont
			{Ahmed}}, \bibinfo {author} {\bibfnamefont {C.}~\bibnamefont {S\'anchez
				Mu\~noz}}, \bibinfo {author} {\bibfnamefont {F.}~\bibnamefont {Nori}},\ and\
		\bibinfo {author} {\bibfnamefont {A.~F.}\ \bibnamefont {Kockum}},\ }\bibfield
	{title} {\bibinfo {title} {Quantum state tomography with conditional
			generative adversarial networks},\ }\href
	{https://doi.org/10.1103/PhysRevLett.127.140502} {\bibfield  {journal}
		{\bibinfo  {journal} {Phys. Rev. Lett.}\ }\textbf {\bibinfo {volume} {127}},\
		\bibinfo {pages} {140502} (\bibinfo {year} {2021})}\BibitemShut {NoStop}%
	\bibitem [{\citenamefont {Quek}\ \emph {et~al.}(2021)\citenamefont {Quek},
		\citenamefont {Fort},\ and\ \citenamefont {Ng}}]{Quek2021}%
	\BibitemOpen
	\bibfield  {author} {\bibinfo {author} {\bibfnamefont {Y.}~\bibnamefont
			{Quek}}, \bibinfo {author} {\bibfnamefont {S.}~\bibnamefont {Fort}},\ and\
		\bibinfo {author} {\bibfnamefont {H.~K.}\ \bibnamefont {Ng}},\ }\bibfield
	{title} {\bibinfo {title} {Adaptive quantum state tomography with neural
			networks},\ }\href {https://doi.org/10.1038/s41534-021-00436-9} {\bibfield
		{journal} {\bibinfo  {journal} {npj Quantum Information}\ }\textbf {\bibinfo
			{volume} {7}},\ \bibinfo {pages} {105} (\bibinfo {year} {2021})}\BibitemShut
	{NoStop}%
	\bibitem [{\citenamefont {Koutn\'y}\ \emph {et~al.}(2022)\citenamefont
		{Koutn\'y}, \citenamefont {Motka}, \citenamefont {Hradil}, \citenamefont
		{\ifmmode \check{R}\else \v{R}\fi{}eh\'a\ifmmode~\check{c}\else
			\v{c}\fi{}ek},\ and\ \citenamefont {S\'anchez-Soto}}]{Kotuny2022}%
	\BibitemOpen
	\bibfield  {author} {\bibinfo {author} {\bibfnamefont {D.}~\bibnamefont
			{Koutn\'y}}, \bibinfo {author} {\bibfnamefont {L.}~\bibnamefont {Motka}},
		\bibinfo {author} {\bibfnamefont {Z.~c.~v.}\ \bibnamefont {Hradil}}, \bibinfo
		{author} {\bibfnamefont {J.}~\bibnamefont {\ifmmode \check{R}\else
				\v{R}\fi{}eh\'a\ifmmode~\check{c}\else \v{c}\fi{}ek}},\ and\ \bibinfo
		{author} {\bibfnamefont {L.~L.}\ \bibnamefont {S\'anchez-Soto}},\ }\bibfield
	{title} {\bibinfo {title} {Neural-network quantum state tomography},\ }\href
	{https://doi.org/10.1103/PhysRevA.106.012409} {\bibfield  {journal} {\bibinfo
			{journal} {Phys. Rev. A}\ }\textbf {\bibinfo {volume} {106}},\ \bibinfo
		{pages} {012409} (\bibinfo {year} {2022})}\BibitemShut {NoStop}%
	\bibitem [{\citenamefont {Schmale}\ \emph {et~al.}(2022)\citenamefont
		{Schmale}, \citenamefont {Reh},\ and\ \citenamefont
		{G{\"a}rttner}}]{Schmale2022}%
	\BibitemOpen
	\bibfield  {author} {\bibinfo {author} {\bibfnamefont {T.}~\bibnamefont
			{Schmale}}, \bibinfo {author} {\bibfnamefont {M.}~\bibnamefont {Reh}},\ and\
		\bibinfo {author} {\bibfnamefont {M.}~\bibnamefont {G{\"a}rttner}},\
	}\bibfield  {title} {\bibinfo {title} {Efficient quantum state tomography
			with convolutional neural networks},\ }\href
	{https://doi.org/10.1038/s41534-022-00621-4} {\bibfield  {journal} {\bibinfo
			{journal} {npj Quantum Information}\ }\textbf {\bibinfo {volume} {8}},\
		\bibinfo {pages} {115} (\bibinfo {year} {2022})}\BibitemShut {NoStop}%
	\bibitem [{\citenamefont {Ma}\ \emph {et~al.}(2024)\citenamefont {Ma},
		\citenamefont {Dong}, \citenamefont {Petersen}, \citenamefont {Huang},\ and\
		\citenamefont {Xiang}}]{Ma2024}%
	\BibitemOpen
	\bibfield  {author} {\bibinfo {author} {\bibfnamefont {H.}~\bibnamefont
			{Ma}}, \bibinfo {author} {\bibfnamefont {D.}~\bibnamefont {Dong}}, \bibinfo
		{author} {\bibfnamefont {I.~R.}\ \bibnamefont {Petersen}}, \bibinfo {author}
		{\bibfnamefont {C.-J.}\ \bibnamefont {Huang}},\ and\ \bibinfo {author}
		{\bibfnamefont {G.-Y.}\ \bibnamefont {Xiang}},\ }\bibfield  {title} {\bibinfo
		{title} {Neural networks for quantum state tomography with constrained
			measurements},\ }\href {https://doi.org/10.1007/s11128-024-04522-7}
	{\bibfield  {journal} {\bibinfo  {journal} {Quantum Information Processing}\
		}\textbf {\bibinfo {volume} {23}},\ \bibinfo {pages} {317} (\bibinfo {year}
		{2024})}\BibitemShut {NoStop}%
	\bibitem [{\citenamefont {Krawczyk}\ \emph {et~al.}(2024)\citenamefont
		{Krawczyk}, \citenamefont {Paw\l{}owski}, \citenamefont
		{Ma\ifmmode~\acute{s}\else \'{s}\fi{}ka},\ and\ \citenamefont
		{Roszak}}]{Krawczyk2024}%
	\BibitemOpen
	\bibfield  {author} {\bibinfo {author} {\bibfnamefont {M.}~\bibnamefont
			{Krawczyk}}, \bibinfo {author} {\bibfnamefont {J.}~\bibnamefont
			{Paw\l{}owski}}, \bibinfo {author} {\bibfnamefont {M.~M.}\ \bibnamefont
			{Ma\ifmmode~\acute{s}\else \'{s}\fi{}ka}},\ and\ \bibinfo {author}
		{\bibfnamefont {K.}~\bibnamefont {Roszak}},\ }\bibfield  {title} {\bibinfo
		{title} {Data-driven criteria for quantum correlations},\ }\href
	{https://doi.org/10.1103/PhysRevA.109.022405} {\bibfield  {journal} {\bibinfo
			{journal} {Phys. Rev. A}\ }\textbf {\bibinfo {volume} {109}},\ \bibinfo
		{pages} {022405} (\bibinfo {year} {2024})}\BibitemShut {NoStop}%
	\bibitem [{\citenamefont {Paw\l{}owski}\ and\ \citenamefont
		{Krawczyk}(2024)}]{Pawlowski2024}%
	\BibitemOpen
	\bibfield  {author} {\bibinfo {author} {\bibfnamefont {J.}~\bibnamefont
			{Paw\l{}owski}}\ and\ \bibinfo {author} {\bibfnamefont {M.}~\bibnamefont
			{Krawczyk}},\ }\bibfield  {title} {\bibinfo {title} {Identification of
			quantum entanglement with siamese convolutional neural networks and
			semisupervised learning},\ }\href
	{https://doi.org/10.1103/PhysRevApplied.22.014068} {\bibfield  {journal}
		{\bibinfo  {journal} {Phys. Rev. Appl.}\ }\textbf {\bibinfo {volume} {22}},\
		\bibinfo {pages} {014068} (\bibinfo {year} {2024})}\BibitemShut {NoStop}%
	\bibitem [{\citenamefont {Taghadomi}\ \emph {et~al.}(2025)\citenamefont
		{Taghadomi}, \citenamefont {Mani}, \citenamefont {Fahim},\ and\ \citenamefont
		{Bakouei}}]{Taghadomi2024}%
	\BibitemOpen
	\bibfield  {author} {\bibinfo {author} {\bibfnamefont {N.}~\bibnamefont
			{Taghadomi}}, \bibinfo {author} {\bibfnamefont {A.}~\bibnamefont {Mani}},
		\bibinfo {author} {\bibfnamefont {A.}~\bibnamefont {Fahim}},\ and\ \bibinfo
		{author} {\bibfnamefont {A.}~\bibnamefont {Bakouei}},\ }\bibfield  {title}
	{\bibinfo {title} {Effective detection of quantum discord by using
			convolutional neural networks},\ }\href
	{https://doi.org/10.1007/s42484-025-00267-3} {\bibfield  {journal} {\bibinfo
			{journal} {Quantum Machine Intelligence}\ }\textbf {\bibinfo {volume} {7}},\
		\bibinfo {pages} {40} (\bibinfo {year} {2025})}\BibitemShut {NoStop}%
	\bibitem [{\citenamefont {Chen}\ \emph {et~al.}(2021)\citenamefont {Chen},
		\citenamefont {Pan}, \citenamefont {Zhang},\ and\ \citenamefont
		{Cheng}}]{Chen2022}%
	\BibitemOpen
	\bibfield  {author} {\bibinfo {author} {\bibfnamefont {Y.}~\bibnamefont
			{Chen}}, \bibinfo {author} {\bibfnamefont {Y.}~\bibnamefont {Pan}}, \bibinfo
		{author} {\bibfnamefont {G.}~\bibnamefont {Zhang}},\ and\ \bibinfo {author}
		{\bibfnamefont {S.}~\bibnamefont {Cheng}},\ }\bibfield  {title} {\bibinfo
		{title} {Detecting quantum entanglement with unsupervised learning},\ }\href
	{https://doi.org/10.1088/2058-9565/ac310f} {\bibfield  {journal} {\bibinfo
			{journal} {Quantum Science and Technology}\ }\textbf {\bibinfo {volume}
			{7}},\ \bibinfo {pages} {015005} (\bibinfo {year} {2021})}\BibitemShut
	{NoStop}%
	\bibitem [{\citenamefont {Asif}\ \emph {et~al.}(2023)\citenamefont {Asif},
		\citenamefont {Khalid}, \citenamefont {Khan}, \citenamefont {Duong},\ and\
		\citenamefont {Shin}}]{Asif2023}%
	\BibitemOpen
	\bibfield  {author} {\bibinfo {author} {\bibfnamefont {N.}~\bibnamefont
			{Asif}}, \bibinfo {author} {\bibfnamefont {U.}~\bibnamefont {Khalid}},
		\bibinfo {author} {\bibfnamefont {A.}~\bibnamefont {Khan}}, \bibinfo {author}
		{\bibfnamefont {T.~Q.}\ \bibnamefont {Duong}},\ and\ \bibinfo {author}
		{\bibfnamefont {H.}~\bibnamefont {Shin}},\ }\bibfield  {title} {\bibinfo
		{title} {Entanglement detection with artificial neural networks},\ }\href
	{https://doi.org/10.1038/s41598-023-28745-3} {\bibfield  {journal} {\bibinfo
			{journal} {Scientific Reports}\ }\textbf {\bibinfo {volume} {13}},\ \bibinfo
		{pages} {1562} (\bibinfo {year} {2023})}\BibitemShut {NoStop}%
	\bibitem [{\citenamefont {Ure{\~n}a}\ \emph {et~al.}(2024)\citenamefont
		{Ure{\~n}a}, \citenamefont {Sojo}, \citenamefont {Bermejo-Vega},\ and\
		\citenamefont {Manzano}}]{urena2024}%
	\BibitemOpen
	\bibfield  {author} {\bibinfo {author} {\bibfnamefont {J.}~\bibnamefont
			{Ure{\~n}a}}, \bibinfo {author} {\bibfnamefont {A.}~\bibnamefont {Sojo}},
		\bibinfo {author} {\bibfnamefont {J.}~\bibnamefont {Bermejo-Vega}},\ and\
		\bibinfo {author} {\bibfnamefont {D.}~\bibnamefont {Manzano}},\ }\bibfield
	{title} {\bibinfo {title} {Entanglement detection with classical deep neural
			networks},\ }\href {https://doi.org/10.1038/s41598-024-68213-0} {\bibfield
		{journal} {\bibinfo  {journal} {Scientific Reports}\ }\textbf {\bibinfo
			{volume} {14}},\ \bibinfo {pages} {18109} (\bibinfo {year}
		{2024})}\BibitemShut {NoStop}%
	\bibitem [{\citenamefont {Lu}\ \emph {et~al.}(2018)\citenamefont {Lu},
		\citenamefont {Huang}, \citenamefont {Li}, \citenamefont {Li}, \citenamefont
		{Chen}, \citenamefont {Lu}, \citenamefont {Ji}, \citenamefont {Shen},
		\citenamefont {Zhou},\ and\ \citenamefont {Zeng}}]{Lu2018}%
	\BibitemOpen
	\bibfield  {author} {\bibinfo {author} {\bibfnamefont {S.}~\bibnamefont
			{Lu}}, \bibinfo {author} {\bibfnamefont {S.}~\bibnamefont {Huang}}, \bibinfo
		{author} {\bibfnamefont {K.}~\bibnamefont {Li}}, \bibinfo {author}
		{\bibfnamefont {J.}~\bibnamefont {Li}}, \bibinfo {author} {\bibfnamefont
			{J.}~\bibnamefont {Chen}}, \bibinfo {author} {\bibfnamefont {D.}~\bibnamefont
			{Lu}}, \bibinfo {author} {\bibfnamefont {Z.}~\bibnamefont {Ji}}, \bibinfo
		{author} {\bibfnamefont {Y.}~\bibnamefont {Shen}}, \bibinfo {author}
		{\bibfnamefont {D.}~\bibnamefont {Zhou}},\ and\ \bibinfo {author}
		{\bibfnamefont {B.}~\bibnamefont {Zeng}},\ }\bibfield  {title} {\bibinfo
		{title} {Separability-entanglement classifier via machine learning},\ }\href
	{https://doi.org/10.1103/PhysRevA.98.012315} {\bibfield  {journal} {\bibinfo
			{journal} {Phys. Rev. A}\ }\textbf {\bibinfo {volume} {98}},\ \bibinfo
		{pages} {012315} (\bibinfo {year} {2018})}\BibitemShut {NoStop}%
	\bibitem [{\citenamefont {Hiesmayr}(2021)}]{Hiesmayr2021}%
	\BibitemOpen
	\bibfield  {author} {\bibinfo {author} {\bibfnamefont {B.~C.}\ \bibnamefont
			{Hiesmayr}},\ }\bibfield  {title} {\bibinfo {title} {Free versus bound
			entanglement, a np-hard problem tackled by machine learning},\ }\href
	{https://doi.org/10.1038/s41598-021-98523-6} {\bibfield  {journal} {\bibinfo
			{journal} {Scientific Reports}\ }\textbf {\bibinfo {volume} {11}},\ \bibinfo
		{pages} {19739} (\bibinfo {year} {2021})}\BibitemShut {NoStop}%
	\bibitem [{\citenamefont {Goes}\ \emph {et~al.}(2021)\citenamefont {Goes},
		\citenamefont {Canabarro}, \citenamefont {Duzzioni},\ and\ \citenamefont
		{Maciel}}]{Goes2024}%
	\BibitemOpen
	\bibfield  {author} {\bibinfo {author} {\bibfnamefont {C.~B.~D.}\
			\bibnamefont {Goes}}, \bibinfo {author} {\bibfnamefont {A.}~\bibnamefont
			{Canabarro}}, \bibinfo {author} {\bibfnamefont {E.~I.}\ \bibnamefont
			{Duzzioni}},\ and\ \bibinfo {author} {\bibfnamefont {T.~O.}\ \bibnamefont
			{Maciel}},\ }\bibfield  {title} {\bibinfo {title} {Automated machine learning
			can classify bound entangled states with tomograms},\ }\href
	{https://doi.org/10.1007/s11128-021-03037-9} {\bibfield  {journal} {\bibinfo
			{journal} {Quantum Information Processing}\ }\textbf {\bibinfo {volume}
			{20}},\ \bibinfo {pages} {99} (\bibinfo {year} {2021})}\BibitemShut {NoStop}%
	\bibitem [{\citenamefont {Wang}(2024)}]{wang2024}%
	\BibitemOpen
	\bibfield  {author} {\bibinfo {author} {\bibfnamefont {B.}~\bibnamefont
			{Wang}},\ }\href {https://arxiv.org/abs/1709.03617} {\bibinfo {title}
		{Learning to detect entanglement}} (\bibinfo {year} {2024}),\ \Eprint
	{https://arxiv.org/abs/1709.03617} {arXiv:1709.03617 [quant-ph]} \BibitemShut
	{NoStop}%
	\bibitem [{\citenamefont {Ganaie}\ \emph {et~al.}(2022)\citenamefont {Ganaie},
		\citenamefont {Hu}, \citenamefont {Malik}, \citenamefont {Tanveer},\ and\
		\citenamefont {Suganthan}}]{Ganaie2022}%
	\BibitemOpen
	\bibfield  {author} {\bibinfo {author} {\bibfnamefont {M.}~\bibnamefont
			{Ganaie}}, \bibinfo {author} {\bibfnamefont {M.}~\bibnamefont {Hu}}, \bibinfo
		{author} {\bibfnamefont {A.}~\bibnamefont {Malik}}, \bibinfo {author}
		{\bibfnamefont {M.}~\bibnamefont {Tanveer}},\ and\ \bibinfo {author}
		{\bibfnamefont {P.}~\bibnamefont {Suganthan}},\ }\bibfield  {title} {\bibinfo
		{title} {Ensemble deep learning: A review},\ }\href
	{https://doi.org/https://doi.org/10.1016/j.engappai.2022.105151} {\bibfield
		{journal} {\bibinfo  {journal} {Eng. Appl. Art. Intel.}\ }\textbf {\bibinfo
			{volume} {115}},\ \bibinfo {pages} {105151} (\bibinfo {year}
		{2022})}\BibitemShut {NoStop}%
	\bibitem [{\citenamefont {Breiman}(2001)}]{Breiman2001}%
	\BibitemOpen
	\bibfield  {author} {\bibinfo {author} {\bibfnamefont {L.}~\bibnamefont
			{Breiman}},\ }\bibfield  {title} {\bibinfo {title} {Random forests},\ }\href
	{https://doi.org/10.1023/A:1010933404324} {\bibfield  {journal} {\bibinfo
			{journal} {Machine Learning}\ }\textbf {\bibinfo {volume} {45}},\ \bibinfo
		{pages} {5} (\bibinfo {year} {2001})}\BibitemShut {NoStop}%
	\bibitem [{\citenamefont {F{\"u}rnkranz}(2010)}]{Furnkranz2010}%
	\BibitemOpen
	\bibfield  {author} {\bibinfo {author} {\bibfnamefont {J.}~\bibnamefont
			{F{\"u}rnkranz}},\ }\bibinfo {title} {Decision tree},\ in\ \href
	{https://doi.org/10.1007/978-0-387-30164-8_204} {\emph {\bibinfo {booktitle}
			{Encyclopedia of Machine Learning}}},\ \bibinfo {editor} {edited by\ \bibinfo
		{editor} {\bibfnamefont {C.}~\bibnamefont {Sammut}}\ and\ \bibinfo {editor}
		{\bibfnamefont {G.~I.}\ \bibnamefont {Webb}}}\ (\bibinfo  {publisher}
	{Springer US},\ \bibinfo {address} {Boston, MA},\ \bibinfo {year} {2010})\
	pp.\ \bibinfo {pages} {263--267}\BibitemShut {NoStop}%
	\bibitem [{\citenamefont {Breiman}\ \emph {et~al.}(2017)\citenamefont
		{Breiman}, \citenamefont {Friedman}, \citenamefont {Olshen},\ and\
		\citenamefont {Stone}}]{breiman2017classification}%
	\BibitemOpen
	\bibfield  {author} {\bibinfo {author} {\bibfnamefont {L.}~\bibnamefont
			{Breiman}}, \bibinfo {author} {\bibfnamefont {J.}~\bibnamefont {Friedman}},
		\bibinfo {author} {\bibfnamefont {R.~A.}\ \bibnamefont {Olshen}},\ and\
		\bibinfo {author} {\bibfnamefont {C.~J.}\ \bibnamefont {Stone}},\ }\href@noop
	{} {\emph {\bibinfo {title} {Classification and regression trees}}}\
	(\bibinfo  {publisher} {Routledge},\ \bibinfo {year} {2017})\BibitemShut
	{NoStop}%
	\bibitem [{\citenamefont {Breiman}(1996)}]{Breiman1996}%
	\BibitemOpen
	\bibfield  {author} {\bibinfo {author} {\bibfnamefont {L.}~\bibnamefont
			{Breiman}},\ }\bibfield  {title} {\bibinfo {title} {Bagging predictors},\
	}\href {https://doi.org/10.1007/BF00058655} {\bibfield  {journal} {\bibinfo
			{journal} {Machine Learning}\ }\textbf {\bibinfo {volume} {24}},\ \bibinfo
		{pages} {123} (\bibinfo {year} {1996})}\BibitemShut {NoStop}%
	\bibitem [{\citenamefont {LeCun}\ \emph {et~al.}(2015)\citenamefont {LeCun},
		\citenamefont {Bengio},\ and\ \citenamefont {Hinton}}]{LeCun2015}%
	\BibitemOpen
	\bibfield  {author} {\bibinfo {author} {\bibfnamefont {Y.}~\bibnamefont
			{LeCun}}, \bibinfo {author} {\bibfnamefont {Y.}~\bibnamefont {Bengio}},\ and\
		\bibinfo {author} {\bibfnamefont {G.}~\bibnamefont {Hinton}},\ }\bibfield
	{title} {\bibinfo {title} {Deep learning},\ }\href
	{https://doi.org/10.1038/nature14539} {\bibfield  {journal} {\bibinfo
			{journal} {Nature}\ }\textbf {\bibinfo {volume} {521}},\ \bibinfo {pages}
		{436} (\bibinfo {year} {2015})}\BibitemShut {NoStop}%
	\bibitem [{\citenamefont {Ronneberger}\ \emph {et~al.}(2015)\citenamefont
		{Ronneberger}, \citenamefont {Fischer},\ and\ \citenamefont {Brox}}]{unet}%
	\BibitemOpen
	\bibfield  {author} {\bibinfo {author} {\bibfnamefont {O.}~\bibnamefont
			{Ronneberger}}, \bibinfo {author} {\bibfnamefont {P.}~\bibnamefont
			{Fischer}},\ and\ \bibinfo {author} {\bibfnamefont {T.}~\bibnamefont
			{Brox}},\ }\bibfield  {title} {\bibinfo {title} {U-net: Convolutional
			networks for biomedical image segmentation},\ }in\ \href
	{https://doi.org/https://doi.org/10.1007/978-3-319-24574-4_28} {\emph
		{\bibinfo {booktitle} {Medical Image Computing and Computer-Assisted
				Intervention -- MICCAI 2015}}},\ \bibinfo {editor} {edited by\ \bibinfo
		{editor} {\bibfnamefont {N.}~\bibnamefont {Navab}}, \bibinfo {editor}
		{\bibfnamefont {J.}~\bibnamefont {Hornegger}}, \bibinfo {editor}
		{\bibfnamefont {W.~M.}\ \bibnamefont {Wells}},\ and\ \bibinfo {editor}
		{\bibfnamefont {A.~F.}\ \bibnamefont {Frangi}}}\ (\bibinfo  {publisher}
	{Springer International Publishing},\ \bibinfo {address} {Cham},\ \bibinfo
	{year} {2015})\ pp.\ \bibinfo {pages} {234--241}\BibitemShut {NoStop}%
	\bibitem [{\citenamefont {Vaswani}\ \emph {et~al.}(2017)\citenamefont
		{Vaswani}, \citenamefont {Shazeer}, \citenamefont {Parmar}, \citenamefont
		{Uszkoreit}, \citenamefont {Jones}, \citenamefont {Gomez}, \citenamefont
		{Kaiser},\ and\ \citenamefont {Polosukhin}}]{vaswani2017}%
	\BibitemOpen
	\bibfield  {author} {\bibinfo {author} {\bibfnamefont {A.}~\bibnamefont
			{Vaswani}}, \bibinfo {author} {\bibfnamefont {N.}~\bibnamefont {Shazeer}},
		\bibinfo {author} {\bibfnamefont {N.}~\bibnamefont {Parmar}}, \bibinfo
		{author} {\bibfnamefont {J.}~\bibnamefont {Uszkoreit}}, \bibinfo {author}
		{\bibfnamefont {L.}~\bibnamefont {Jones}}, \bibinfo {author} {\bibfnamefont
			{A.~N.}\ \bibnamefont {Gomez}}, \bibinfo {author} {\bibfnamefont
			{{\L}.}~\bibnamefont {Kaiser}},\ and\ \bibinfo {author} {\bibfnamefont
			{I.}~\bibnamefont {Polosukhin}},\ }\bibfield  {title} {\bibinfo {title}
		{Attention is all you need},\ }in\ \href {https://arxiv.org/abs/1706.03762}
	{\emph {\bibinfo {booktitle} {Advances in Neural Information Processing
				Systems}}}\ (\bibinfo {year} {2017})\ pp.\ \bibinfo {pages}
	{5998--6008}\BibitemShut {NoStop}%
	\bibitem [{\citenamefont {Gibney}\ and\ \citenamefont
		{Castelvecchi}(2024)}]{Gibney2024}%
	\BibitemOpen
	\bibfield  {author} {\bibinfo {author} {\bibfnamefont {E.}~\bibnamefont
			{Gibney}}\ and\ \bibinfo {author} {\bibfnamefont {D.}~\bibnamefont
			{Castelvecchi}},\ }\bibfield  {title} {\bibinfo {title} {Physics nobel
			scooped by machine-learning pioneers},\ }\href
	{https://www.nature.com/articles/d41586-024-03213-8} {\bibfield  {journal}
		{\bibinfo  {journal} {Nature}\ }\textbf {\bibinfo {volume} {634}},\ \bibinfo
		{pages} {523} (\bibinfo {year} {2024})}\BibitemShut {NoStop}%
	\bibitem [{\citenamefont {Belkin}\ \emph {et~al.}(2019)\citenamefont {Belkin},
		\citenamefont {Hsu}, \citenamefont {Ma},\ and\ \citenamefont
		{Mandal}}]{belkin2021}%
	\BibitemOpen
	\bibfield  {author} {\bibinfo {author} {\bibfnamefont {M.}~\bibnamefont
			{Belkin}}, \bibinfo {author} {\bibfnamefont {D.}~\bibnamefont {Hsu}},
		\bibinfo {author} {\bibfnamefont {S.}~\bibnamefont {Ma}},\ and\ \bibinfo
		{author} {\bibfnamefont {S.}~\bibnamefont {Mandal}},\ }\bibfield  {title}
	{\bibinfo {title} {Reconciling modern machine-learning practice and the
			classical bias–variance trade-off},\ }\href
	{https://doi.org/10.1073/pnas.1903070116} {\bibfield  {journal} {\bibinfo
			{journal} {Proceedings of the National Academy of Sciences}\ }\textbf
		{\bibinfo {volume} {116}},\ \bibinfo {pages} {15849} (\bibinfo {year}
		{2019})},\ \Eprint
	{https://arxiv.org/abs/https://www.pnas.org/doi/pdf/10.1073/pnas.1903070116}
	{https://www.pnas.org/doi/pdf/10.1073/pnas.1903070116} \BibitemShut {NoStop}%
	\bibitem [{\citenamefont {Longo}\ \emph {et~al.}(2024)\citenamefont {Longo},
		\citenamefont {Brcic}, \citenamefont {Cabitza}, \citenamefont {Choi},
		\citenamefont {Confalonieri}, \citenamefont {Ser}, \citenamefont {Guidotti},
		\citenamefont {Hayashi}, \citenamefont {Herrera}, \citenamefont {Holzinger},
		\citenamefont {Jiang}, \citenamefont {Khosravi}, \citenamefont {Lecue},
		\citenamefont {Malgieri}, \citenamefont {Páez}, \citenamefont {Samek},
		\citenamefont {Schneider}, \citenamefont {Speith},\ and\ \citenamefont
		{Stumpf}}]{Longo2024}%
	\BibitemOpen
	\bibfield  {author} {\bibinfo {author} {\bibfnamefont {L.}~\bibnamefont
			{Longo}}, \bibinfo {author} {\bibfnamefont {M.}~\bibnamefont {Brcic}},
		\bibinfo {author} {\bibfnamefont {F.}~\bibnamefont {Cabitza}}, \bibinfo
		{author} {\bibfnamefont {J.}~\bibnamefont {Choi}}, \bibinfo {author}
		{\bibfnamefont {R.}~\bibnamefont {Confalonieri}}, \bibinfo {author}
		{\bibfnamefont {J.~D.}\ \bibnamefont {Ser}}, \bibinfo {author} {\bibfnamefont
			{R.}~\bibnamefont {Guidotti}}, \bibinfo {author} {\bibfnamefont
			{Y.}~\bibnamefont {Hayashi}}, \bibinfo {author} {\bibfnamefont
			{F.}~\bibnamefont {Herrera}}, \bibinfo {author} {\bibfnamefont
			{A.}~\bibnamefont {Holzinger}}, \bibinfo {author} {\bibfnamefont
			{R.}~\bibnamefont {Jiang}}, \bibinfo {author} {\bibfnamefont
			{H.}~\bibnamefont {Khosravi}}, \bibinfo {author} {\bibfnamefont
			{F.}~\bibnamefont {Lecue}}, \bibinfo {author} {\bibfnamefont
			{G.}~\bibnamefont {Malgieri}}, \bibinfo {author} {\bibfnamefont
			{A.}~\bibnamefont {Páez}}, \bibinfo {author} {\bibfnamefont
			{W.}~\bibnamefont {Samek}}, \bibinfo {author} {\bibfnamefont
			{J.}~\bibnamefont {Schneider}}, \bibinfo {author} {\bibfnamefont
			{T.}~\bibnamefont {Speith}},\ and\ \bibinfo {author} {\bibfnamefont
			{S.}~\bibnamefont {Stumpf}},\ }\bibfield  {title} {\bibinfo {title}
		{Explainable artificial intelligence (xai) 2.0: A manifesto of open
			challenges and interdisciplinary research directions},\ }\href
	{https://doi.org/https://doi.org/10.1016/j.inffus.2024.102301} {\bibfield
		{journal} {\bibinfo  {journal} {Information Fusion}\ }\textbf {\bibinfo
			{volume} {106}},\ \bibinfo {pages} {102301} (\bibinfo {year}
		{2024})}\BibitemShut {NoStop}%
	\bibitem [{\citenamefont {Wootters}(1998)}]{Wootters1998}%
	\BibitemOpen
	\bibfield  {author} {\bibinfo {author} {\bibfnamefont {W.~K.}\ \bibnamefont
			{Wootters}},\ }\bibfield  {title} {\bibinfo {title} {Entanglement of
			formation of an arbitrary state of two qubits},\ }\href
	{https://doi.org/10.1103/PhysRevLett.80.2245} {\bibfield  {journal} {\bibinfo
			{journal} {Phys. Rev. Lett.}\ }\textbf {\bibinfo {volume} {80}},\ \bibinfo
		{pages} {2245} (\bibinfo {year} {1998})}\BibitemShut {NoStop}%
	\bibitem [{\citenamefont {Horodecki}(1997)}]{horodecki97}%
	\BibitemOpen
	\bibfield  {author} {\bibinfo {author} {\bibfnamefont {P.}~\bibnamefont
			{Horodecki}},\ }\bibfield  {title} {\bibinfo {title} {Separability criterion
			and inseparable mixed states with positive partial transposition},\ }\href
	{https://doi.org/https://doi.org/10.1016/S0375-9601(97)00416-7} {\bibfield
		{journal} {\bibinfo  {journal} {Physics Letters A}\ }\textbf {\bibinfo
			{volume} {232}},\ \bibinfo {pages} {333} (\bibinfo {year}
		{1997})}\BibitemShut {NoStop}%
	\bibitem [{\citenamefont {Horodecki}\ \emph {et~al.}(1998)\citenamefont
		{Horodecki}, \citenamefont {Horodecki},\ and\ \citenamefont
		{Horodecki}}]{horodecki98}%
	\BibitemOpen
	\bibfield  {author} {\bibinfo {author} {\bibfnamefont {M.}~\bibnamefont
			{Horodecki}}, \bibinfo {author} {\bibfnamefont {P.}~\bibnamefont
			{Horodecki}},\ and\ \bibinfo {author} {\bibfnamefont {R.}~\bibnamefont
			{Horodecki}},\ }\bibfield  {title} {\bibinfo {title} {Mixed-state
			entanglement and distillation: Is there a ``bound'' entanglement in
			nature?},\ }\href {https://doi.org/10.1103/PhysRevLett.80.5239} {\bibfield
		{journal} {\bibinfo  {journal} {Phys. Rev. Lett.}\ }\textbf {\bibinfo
			{volume} {80}},\ \bibinfo {pages} {5239} (\bibinfo {year}
		{1998})}\BibitemShut {NoStop}%
	\bibitem [{\citenamefont {Yu}\ and\ \citenamefont {Eberly}(2007)}]{yu07}%
	\BibitemOpen
	\bibfield  {author} {\bibinfo {author} {\bibfnamefont {T.}~\bibnamefont
			{Yu}}\ and\ \bibinfo {author} {\bibfnamefont {J.~H.}\ \bibnamefont
			{Eberly}},\ }\bibfield  {title} {\bibinfo {title} {Evolution from
			entanglement to decoherence of bipartite mixed "x" states},\ }\href
	{https://dl.acm.org/doi/10.5555/2011832.2011835} {\bibfield  {journal}
		{\bibinfo  {journal} {Quantum Information and Computation}\ }\textbf
		{\bibinfo {volume} {7}},\ \bibinfo {pages} {459} (\bibinfo {year}
		{2007})}\BibitemShut {NoStop}%
	\bibitem [{\citenamefont {Mendonça}\ \emph {et~al.}(2014)\citenamefont
		{Mendonça}, \citenamefont {Marchiolli},\ and\ \citenamefont
		{Galetti}}]{MENDONCA201479}%
	\BibitemOpen
	\bibfield  {author} {\bibinfo {author} {\bibfnamefont {P.~E.}\ \bibnamefont
			{Mendonça}}, \bibinfo {author} {\bibfnamefont {M.~A.}\ \bibnamefont
			{Marchiolli}},\ and\ \bibinfo {author} {\bibfnamefont {D.}~\bibnamefont
			{Galetti}},\ }\bibfield  {title} {\bibinfo {title} {Entanglement universality
			of two-qubit x-states},\ }\href
	{https://doi.org/https://doi.org/10.1016/j.aop.2014.08.017} {\bibfield
		{journal} {\bibinfo  {journal} {Annals of Physics}\ }\textbf {\bibinfo
			{volume} {351}},\ \bibinfo {pages} {79} (\bibinfo {year} {2014})}\BibitemShut
	{NoStop}%
	\bibitem [{\citenamefont {Mezzadri}(2006)}]{mezzadri2006}%
	\BibitemOpen
	\bibfield  {author} {\bibinfo {author} {\bibfnamefont {F.}~\bibnamefont
			{Mezzadri}},\ }\bibfield  {title} {\bibinfo {title} {How to generate random
			matrices from the classical compact groups},\ }\href
	{http://www.ams.org/notices/200705/index.html} {\bibfield  {journal}
		{\bibinfo  {journal} {Notices of the American Mathematical Society}\ }\textbf
		{\bibinfo {volume} {54}} (\bibinfo {year} {2006})}\BibitemShut {NoStop}%
	\bibitem [{\citenamefont {Schapire}(1990)}]{schapire1990strength}%
	\BibitemOpen
	\bibfield  {author} {\bibinfo {author} {\bibfnamefont {R.~E.}\ \bibnamefont
			{Schapire}},\ }\bibfield  {title} {\bibinfo {title} {The strength of weak
			learnability},\ }\href {https://doi.org/https://doi.org/10.1007/BF00116037}
	{\bibfield  {journal} {\bibinfo  {journal} {Machine learning}\ }\textbf
		{\bibinfo {volume} {5}},\ \bibinfo {pages} {197} (\bibinfo {year}
		{1990})}\BibitemShut {NoStop}%
	\bibitem [{\citenamefont {Breiman}(1998)}]{breiman1998arcing}%
	\BibitemOpen
	\bibfield  {author} {\bibinfo {author} {\bibfnamefont {L.}~\bibnamefont
			{Breiman}},\ }\bibfield  {title} {\bibinfo {title} {Arcing classifier (with
			discussion and a rejoinder by the author)},\ }\href
	{https://doi.org/10.1214/aos/1024691079} {\bibfield  {journal} {\bibinfo
			{journal} {The Annals of Statistics}\ }\textbf {\bibinfo {volume} {26}},\
		\bibinfo {pages} {801} (\bibinfo {year} {1998})}\BibitemShut {NoStop}%
	\bibitem [{\citenamefont {Nair}\ and\ \citenamefont {Hinton}(2010)}]{Nair2010}%
	\BibitemOpen
	\bibfield  {author} {\bibinfo {author} {\bibfnamefont {V.}~\bibnamefont
			{Nair}}\ and\ \bibinfo {author} {\bibfnamefont {G.~E.}\ \bibnamefont
			{Hinton}},\ }\bibfield  {title} {\bibinfo {title} {Rectified linear units
			improve restricted boltzmann machines},\ }in\ \href
	{https://www.cs.toronto.edu/~hinton/absps/reluICML.pdf} {\emph {\bibinfo
			{booktitle} {Proceedings of the 27th International Conference on
				International Conference on Machine Learning}}},\ \bibinfo {series and
		number} {ICML'10}\ (\bibinfo  {publisher} {Omnipress},\ \bibinfo {address}
	{Madison, WI, USA},\ \bibinfo {year} {2010})\ p.\ \bibinfo {pages}
	{807–814}\BibitemShut {NoStop}%
	\bibitem [{\citenamefont {Raghavan}\ \emph {et~al.}(1989)\citenamefont
		{Raghavan}, \citenamefont {Bollmann},\ and\ \citenamefont
		{Jung}}]{Raghavan1989}%
	\BibitemOpen
	\bibfield  {author} {\bibinfo {author} {\bibfnamefont {V.}~\bibnamefont
			{Raghavan}}, \bibinfo {author} {\bibfnamefont {P.}~\bibnamefont {Bollmann}},\
		and\ \bibinfo {author} {\bibfnamefont {G.~S.}\ \bibnamefont {Jung}},\
	}\bibfield  {title} {\bibinfo {title} {A critical investigation of recall and
			precision as measures of retrieval system performance},\ }\href
	{https://doi.org/10.1145/65943.65945} {\bibfield  {journal} {\bibinfo
			{journal} {ACM Trans. Inf. Syst.}\ }\textbf {\bibinfo {volume} {7}},\
		\bibinfo {pages} {205–229} (\bibinfo {year} {1989})}\BibitemShut {NoStop}%
	\bibitem [{\citenamefont {Shapley}(1953)}]{shapley1953}%
	\BibitemOpen
	\bibfield  {author} {\bibinfo {author} {\bibfnamefont {L.~S.}\ \bibnamefont
			{Shapley}},\ }\bibfield  {title} {\bibinfo {title} {A value for n-person
			games},\ }in\ \href
	{https://doi.org/https://doi.org/10.1515/9781400881970-018} {\emph {\bibinfo
			{booktitle} {Contributions to the Theory of Games}}},\ Vol.~\bibinfo {volume}
	{2},\ \bibinfo {editor} {edited by\ \bibinfo {editor} {\bibfnamefont {H.~W.}\
			\bibnamefont {Kuhn}}\ and\ \bibinfo {editor} {\bibfnamefont {A.~W.}\
			\bibnamefont {Tucker}}}\ (\bibinfo  {publisher} {Princeton University
		Press},\ \bibinfo {year} {1953})\ pp.\ \bibinfo {pages}
	{307--317}\BibitemShut {NoStop}%
	\bibitem [{\citenamefont {Lundberg}\ and\ \citenamefont
		{Lee}(2017)}]{Lundberg2017}%
	\BibitemOpen
	\bibfield  {author} {\bibinfo {author} {\bibfnamefont {S.~M.}\ \bibnamefont
			{Lundberg}}\ and\ \bibinfo {author} {\bibfnamefont {S.-I.}\ \bibnamefont
			{Lee}},\ }\bibfield  {title} {\bibinfo {title} {A unified approach to
			interpreting model predictions},\ }in\ \href
	{https://proceedings.neurips.cc/paper_files/paper/2017/file/8a20a8621978632d76c43dfd28b67767-Paper.pdf}
	{\emph {\bibinfo {booktitle} {Proceedings of the 31st International
				Conference on Neural Information Processing Systems}}},\ \bibinfo {series and
		number} {NIPS'17}\ (\bibinfo  {publisher} {Curran Associates Inc.},\ \bibinfo
	{address} {Red Hook, NY, USA},\ \bibinfo {year} {2017})\ p.\ \bibinfo {pages}
	{4768–4777}\BibitemShut {NoStop}%
	\bibitem [{\citenamefont {Jolliffe}(2002)}]{Jolliffe2002}%
	\BibitemOpen
	\bibfield  {author} {\bibinfo {author} {\bibfnamefont {I.~T.}\ \bibnamefont
			{Jolliffe}},\ }\href {https://doi.org/10.1007/b98835} {\emph {\bibinfo
			{title} {Principal Component Analysis}}},\ Springer Series in Statistics\
	(\bibinfo  {publisher} {Springer},\ \bibinfo {address} {New York, NY},\
	\bibinfo {year} {2002})\BibitemShut {NoStop}%
	\bibitem [{\citenamefont {Hinton}\ and\ \citenamefont
		{Roweis}(2003)}]{hinton2003sne}%
	\BibitemOpen
	\bibfield  {author} {\bibinfo {author} {\bibfnamefont {G.~E.}\ \bibnamefont
			{Hinton}}\ and\ \bibinfo {author} {\bibfnamefont {S.~T.}\ \bibnamefont
			{Roweis}},\ }\bibfield  {title} {\bibinfo {title} {Stochastic neighbor
			embedding},\ }in\ \href
	{https://proceedings.neurips.cc/paper_files/paper/2002/file/6150ccc6069bea6b5716254057a194ef-Paper.pdf}
	{\emph {\bibinfo {booktitle} {Advances in Neural Information Processing
				Systems}}},\ Vol.~\bibinfo {volume} {15}\ (\bibinfo  {publisher} {MIT
		Press},\ \bibinfo {year} {2003})\ pp.\ \bibinfo {pages}
	{857--864}\BibitemShut {NoStop}%
	\bibitem [{\citenamefont {Scornet}(2023)}]{scornet2023trees}%
	\BibitemOpen
	\bibfield  {author} {\bibinfo {author} {\bibfnamefont {E.}~\bibnamefont
			{Scornet}},\ }\bibfield  {title} {\bibinfo {title} {Trees, forests, and
			impurity-based variable importance in regression},\ }in\ \href
	{https://doi.org/10.1214/21-AIHP1240} {\emph {\bibinfo {booktitle} {Annales
				de l'Institut Henri Poincare (B) Probabilites et statistiques}}},\
	Vol.~\bibinfo {volume} {59}\ (\bibinfo {organization} {Institut Henri
		Poincar{\'e}},\ \bibinfo {year} {2023})\ pp.\ \bibinfo {pages}
	{21--52}\BibitemShut {NoStop}%
	\bibitem [{\citenamefont {Pedregosa}\ \emph {et~al.}(2011)\citenamefont
		{Pedregosa}, \citenamefont {Varoquaux}, \citenamefont {Gramfort},
		\citenamefont {Michel}, \citenamefont {Thirion}, \citenamefont {Grisel},
		\citenamefont {Blondel}, \citenamefont {Prettenhofer}, \citenamefont {Weiss},
		\citenamefont {Dubourg}, \citenamefont {Vanderplas}, \citenamefont {Passos},
		\citenamefont {Cournapeau}, \citenamefont {Brucher}, \citenamefont {Perrot},\
		and\ \citenamefont {Duchesnay}}]{scikit-learn}%
	\BibitemOpen
	\bibfield  {author} {\bibinfo {author} {\bibfnamefont {F.}~\bibnamefont
			{Pedregosa}}, \bibinfo {author} {\bibfnamefont {G.}~\bibnamefont
			{Varoquaux}}, \bibinfo {author} {\bibfnamefont {A.}~\bibnamefont {Gramfort}},
		\bibinfo {author} {\bibfnamefont {V.}~\bibnamefont {Michel}}, \bibinfo
		{author} {\bibfnamefont {B.}~\bibnamefont {Thirion}}, \bibinfo {author}
		{\bibfnamefont {O.}~\bibnamefont {Grisel}}, \bibinfo {author} {\bibfnamefont
			{M.}~\bibnamefont {Blondel}}, \bibinfo {author} {\bibfnamefont
			{P.}~\bibnamefont {Prettenhofer}}, \bibinfo {author} {\bibfnamefont
			{R.}~\bibnamefont {Weiss}}, \bibinfo {author} {\bibfnamefont
			{V.}~\bibnamefont {Dubourg}}, \bibinfo {author} {\bibfnamefont
			{J.}~\bibnamefont {Vanderplas}}, \bibinfo {author} {\bibfnamefont
			{A.}~\bibnamefont {Passos}}, \bibinfo {author} {\bibfnamefont
			{D.}~\bibnamefont {Cournapeau}}, \bibinfo {author} {\bibfnamefont
			{M.}~\bibnamefont {Brucher}}, \bibinfo {author} {\bibfnamefont
			{M.}~\bibnamefont {Perrot}},\ and\ \bibinfo {author} {\bibfnamefont
			{E.}~\bibnamefont {Duchesnay}},\ }\bibfield  {title} {\bibinfo {title}
		{Scikit-learn: Machine learning in {P}ython},\ }\href
	{http://jmlr.org/papers/v12/pedregosa11a.html} {\bibfield  {journal}
		{\bibinfo  {journal} {Journal of Machine Learning Research}\ }\textbf
		{\bibinfo {volume} {12}},\ \bibinfo {pages} {2825} (\bibinfo {year}
		{2011})}\BibitemShut {NoStop}%
\end{thebibliography}
\end{document}